\documentclass[reprint,superscriptaddress,aps,twocolumn,showpacs]{revtex4-1}
        \usepackage[vcentermath]{youngtab}
\usepackage{amssymb}
\usepackage{amsmath}
\usepackage{graphicx}
\usepackage{multirow}
\usepackage{hyperref}
\usepackage{longtable}

\allowdisplaybreaks

\begin{document}

\title{Complete basis for the pentaquark wave function in a group theory approach}

\author{K. Xu}
\email[]{gxukai1123@gmail.com}
\author{A. Kaewsnod}
\affiliation{School of Physics and Center of Excellence in High Energy Physics and Astrophysics, Suranaree University of Technology, Nakhon Ratchasima 30000, Thailand}
\author{X. Y. Liu}
\affiliation{School of Physics and Center of Excellence in High Energy Physics and Astrophysics, Suranaree University of Technology, Nakhon Ratchasima 30000, Thailand}
\affiliation{School of Mathematics and Physics, Bohai University, Jinzhou 121013, China}
\author{S. Srisuphaphon}
\affiliation{Department of Physics, Faculty of Science, Burapha University, Chonburi 20131, Thailand}
\author{A. Limphirat}
\author{Y. Yan}
\email[]{yupeng@sut.ac.th}
\affiliation{School of Physics and Center of Excellence in High Energy Physics and Astrophysics, Suranaree University of Technology, Nakhon Ratchasima 30000, Thailand}

\date{\today}

\begin{abstract}
\indent   Permutation groups are applied to analyze the symmetries of pentaquark states. All possible quark configurations of the color, flavor, spin and spatial degrees of freedom are worked out in the language of permutation groups, and the corresponding wave functions are constructed systematically in the form of a Yamanouchi basis. The pentaquark spatial wave functions of various symmetries, which are derived in the harmonic-oscillator interaction, are applied as complete bases to evaluate the low-lying light $q^4\overline q$ pentaquark mass of all configurations, where the Cornell-like potential is employed.

\end{abstract}


\maketitle

\section{Introduction}\label{sec:Int}
\indent  In recent decades, hadron physicists have expended great effort hunting for evidence of the multiquark states. Since the arguable state of $\Theta^{+}$(1540) was proposed as the first observed pentaquark, tremendous progress on experimental and theoretical explorations of the multiquark states has been achieved. Disregarding the large number of XYZ tetraquark candidates, in recent years the LHCb Collaboration has reported and confirmed the observation of three narrow pentaquark-like states: $P_c(4312)^{+}$, $P_c(4440)^{+}$, and $P_c(4457)^{+}$. All three pentaquark-like states may have the quark content of $uudc \bar c$ \cite{LHCb1,LHCb2,LHCb3}, but their internal structures as well as quantum numbers are still unclear.

\indent  A group theory approach has been applied to construct the pentaquark wave functions and study the role of pentaquark components in baryons~\cite{Bijker2004,Bijker2019,An2006,An2009,Wang2009,Sorakrai2012,Kai2014,Sorakrai2016}. However, the construction of high-order spatial wave functions of pentaquark states in various permutation symmetries
has been a challenge. In this work we systematically construct in the group theory approach the pentaquark wave function in the form of a Yamanouchi basis, including high-order spatial excitations. All the basic notations have been introduced in previous studies~\cite{Sorakrai2012,Kai2014}, where the $q^3$ and $q^4 \bar q$ wave functions have been derived partly.

\indent  The paper is organized as follows. In Sec.~\ref{sec:GTM} the properties of the characters of the $S_4$ permutation group are applied to work out all the possible configurations of color, flavor, spin, and spatial degrees of freedom of $q^4 \bar q$ states, and the explicit forms of pentaquark wave functions are derived in the form of a Yamanouchi basis for various configurations. The spatial wave functions of various symmetries are constructed to high orders in the harmonic oscillator interaction. In Sec. ~\ref{sec:NC} we apply the spatial wave functions constructed in Sec.~\ref{sec:GTM} as the complete basis, to evaluate the low-lying light pentaquark mass spectrum in the constituent quark model, where a hamiltonian including a Cornell-like potential and one-gluon exchange contribution as hyperfine interaction is employed and all model parameters are predetermined by comparing the theoretical and experimental masses of low-lying $q^3$ baryons. A summary is given in Sec.~\ref{sec:SUM}.
\section{GROUP THEORY Methods}
\label{sec:GTM}
\subsection{Color-spin-flavor wave function}
\indent The construction of $q^4 \bar q$ states follows the rules that a $q^4 \bar q$ state must be a color
singlet and the $q^4 \bar q$ wave function should be
antisymmetric under any permutation between identical quarks.
The permutation symmetry of the four-quark configuration of pentaquark states is characterized by the $S_4$ young tabloids [4], [31], [22], [211], [1111]. Requiring the pentaquark to be a color singlet demands that the color part of the pentaquark wave function must be a $[222]_1$ singlet. Since the color part of the antiquark in pentaquark states is a $[11]_3$ antitriplet, the color wave function of the four-quark configuration must be a $[211]_3$ triplet. Requiring the total wave function of the four quark configuration to be antisymmetric implies that its spatial-spin-flavor part must be a [31] state by conjugation.

\indent The algebraic structure of a multiquark state consists of
the usual spin-flavor and color algebras
$SU_{sf}(6) \otimes SU_{c}(3)$ with
$SU_{sf}(6)= SU_{f}(3) \otimes SU_{s}(2)$. The four-quark state $|q_1\rangle |q_2\rangle |q_3\rangle |q_4\rangle$ forms an $m^4$-dimensional direct product basis of $SU(m)$ ($m=3,\;3,\;2$ for
the color, flavor, and spin), which can be decomposed according to the $S_4$ permutation group.

The orthogonal theorem in group theory leads to the following property for the characters of a group~\cite{Yan2006}:
\begin{eqnarray}
\chi(g)&=&\sum_{\beta=1}^h\,m_\beta \chi^{(\beta)}(g)
\end{eqnarray}
where $g$ are group elements, $\chi(g)$ are the characters of a product (reducible) representation of the group, and $\chi^{(\beta)}(g)$
are the characters of the irreducible representation labeled by $\beta$. From the above equation, one gets
\begin{eqnarray}\label{eqn::orthogonal}
m_\alpha &=&\frac{1}{n}\sum_{g}\chi^{(\alpha)*}(g)\,\chi(g)
\nonumber \\
&=& \sum_{i=1}^{r} \frac{\rho_i}{n} \chi_i^{(\alpha)*} \chi_i
\end{eqnarray}
Here we have used the fact that the finite group $G=\{g\}$ of order $n$ has $r$ conjugacy classes $C_i$ ($i=1,2,...r$) and the numbers of the elements in the conjugacy class $C_i$ is $\rho_i$.
For the $S_4$ group of order $n=24$ there are five conjugacy classes: (e), (ij), (ij)(kl), (ijk), and (ijkl). The character values of all five $C_i$ of the $S_4$ group are presented in Table \ref{tab:ct} for all irreducible representations.

By applying Eq. (\ref{eqn::orthogonal}), one gets all the spatial-spin-flavor configurations and spin-flavor configurations of the $q^4$ cluster of pentaquarks, as shown in Tables \ref{tab:wf0osf} and \ref{tab:wf0sf}, respectively,

\begin{table}[!htbp]
\caption{Character tables of conjugacy classes of $S_4$}
\vspace*{-0.2cm}
\begin{center}\label{tab:ct}
\begin{tabular}{lclclcccccccccc}
\hline
\hline
\\
& $C_i$ && $\rho_i$ && $\chi^{[4]}$ && $\chi^{[31]}$  && $\chi^{[22]}$ && $\chi^{[211]}$ && $\chi^{[1111]}$
 \\
\hline
 \\
& ($e$) &&      1 && 1 && 3 && 2 && 3 && 1
\\& \\
& $(ij)$  &&      6 && 1 && 1  && 0 && -1 && -1
\\&\\
& $(ij)(kl)$ && 3 && 1 && -1  && 2 && -1 && 1
\\&\\
& $(ijk)$ &&    8 && 1 &&  0  && -1 && 0 && 1
\\& \\
& $(ijkl)$ &&   6 && 1 && -1  && 0 && 1&& -1
 \\
\hline
\end{tabular}
\end{center}
\end{table}

\begin{table}[h]
\caption[Spatial-spin-flavor configurations of $q^4$]{Spatial-spin-flavor configurations of $q^4$.}
\begin{center}\label{tab:wf0osf}
  \begin{tabular}{|l|l|}
   \hline\multicolumn{2}{c}{$[31]_{OSF}$}\\
   \hline\multicolumn{1}{l}{$[4]_{O}$}& \multicolumn{1}{l}{$[31]_{SF}$}\\
         \multicolumn{1}{l}{$[1111]_{O}$}& \multicolumn{1}{l}{$[211]_{SF}$}\\
         \multicolumn{1}{l}{$[22]_{O}$}& \multicolumn{1}{l}{$[31]_{SF},[211]_{SF}$}\\
         \multicolumn{1}{l}{$[211]_{O}$}& \multicolumn{1}{l}{$[31]_{SF},[211]_{SF},[22]_{SF},[1111]_{SF}$}\\
         \multicolumn{1}{l}{$[31]_{O}$}& \multicolumn{1}{l}{$[4]_{SF},[31]_{SF},[211]_{SF},[22]_{SF}$}\\
   \hline
\end{tabular}
\end{center}
\end{table}

\begin{table}[!h]
\caption{Spin-flavor configurations of $q^4$.}
\begin{center}\label{tab:wf0sf}
  \begin{tabular}{lll}
    \hline
   \hline\multicolumn{3}{c}{$[4]_{FS}$}\\&\\
          \multicolumn{1}{l}{$[4]_{FS}[22]_{F}[22]_{S}$}&
          \multicolumn{1}{l}{$[4]_{FS}[31]_{F}[31]_{S}$}&
          \multicolumn{1}{l}{$[4]_{FS}[4]_{F}[4]_{S}$}\\
   \hline\multicolumn{3}{c}{$[31]_{FS}$}\\&\\
          \multicolumn{1}{l}{$[31]_{FS}[31]_{F}[22]_{S}$}&
          \multicolumn{1}{l}{$[31]_{FS}[31]_{F}[31]_{S}$}&
          \multicolumn{1}{l}{$[31]_{FS}[31]_{F}[4]_{S}$}\\&\\
          \multicolumn{1}{l}{$[31]_{FS}[211]_{F}[22]_{S}$}&
          \multicolumn{1}{l}{$[31]_{FS}[211]_{F}[31]_{S}$}&
          \multicolumn{1}{l}{$[31]_{FS}[22]_{F}[31]_{S}$}\\&\\
          \multicolumn{1}{l}{$[31]_{FS}[4]_{F}[31]_{S}$}\\
   \hline\multicolumn{3}{c}{$[22]_{FS}$}\\&\\
          \multicolumn{1}{l}{$[22]_{FS}[22]_{F}[22]_{S}$}&
          \multicolumn{1}{l}{$[22]_{FS}[22]_{F}[4]_{S}$}&
          \multicolumn{1}{l}{$[22]_{FS}[4]_{F}[22]_{S}$}\\&\\
          \multicolumn{1}{l}{$[22]_{FS}[211]_{F}[31]_{S}$}&
          \multicolumn{1}{l}{$[22]_{FS}[31]_{F}[31]_{S}$}\\
   \hline\multicolumn{3}{c}{$[211]_{FS}$}\\&\\
          \multicolumn{1}{l}{$[211]_{FS}[211]_{F}[22]_{S}$}&
          \multicolumn{1}{l}{$[211]_{FS}[211]_{F}[31]_{S}$}&
          \multicolumn{1}{l}{$[211]_{FS}[211]_{F}[4]_{S}$}\\&\\
          \multicolumn{1}{l}{$[211]_{FS}[22]_{F}[31]_{S}$}&
          \multicolumn{1}{l}{$[211]_{FS}[31]_{F}[22]_{S}$}&
          \multicolumn{1}{l}{$[211]_{FS}[31]_{F}[31]_{S}$}\\
   \hline\multicolumn{3}{c}{$[1111]_{FS}$}\\&\\
          \multicolumn{1}{l}{$[1111]_{FS}[211]_{F}[31]_{S}$}&
          \multicolumn{1}{l}{$[1111]_{FS}[22]_{F}[22]_{S}$}\\
    \hline
\end{tabular}
\end{center}
\end{table}

\indent The total wave function of the $q^4$ configuration may be written in
the general form
\begin{eqnarray}\label{eq::1}
\Psi_{total}&=&\sum_{i,j=\rho,\lambda,\eta}a_{ij}\; \psi^c_{[211]_i}\psi^{osf}_{[31]_j},
\end{eqnarray}
with
\begin{eqnarray}\label{eq::2}
\psi^{osf}_{[31]_{\rho,\lambda,\eta}} &=& \sum_{i,j=S,A,\rho,\lambda,\eta}b_{ij}\psi^{o}_{[X]_{i}}\psi^{sf}_{[Y]_{j}}, \nonumber \\
\psi^{sf}_{[Y]} &=& \sum_{i,j=S,A,\rho,\lambda,\eta}c_{ij}\psi^{s}_{[x]_{i}}\psi^{f}_{[y]_{j}},
\end{eqnarray}
where $\psi^c$, $\psi^{osf}$, $\psi^{sf}$, $\psi^{s}$, and $\psi^{f}$ are respectively the color, spatial-spin-flavor, spin-flavor, spin, and flavor parts of the $q^4$ cluster.
$S,\;A,\;\rho,\;\lambda$, and $\eta$ stand for fully symmetric, fully antisymmetric, $\rho$-type, $\lambda$-type, and $\eta$-type functions.
The coefficients in Eqs. (\ref{eq::1}) and (\ref{eq::2}) can be determined by enacting the permutations $(12)$, $(23)$ and $(34)$ of the $S_4$ group on both sides of the general wave functions.
The fully antisymmetric wave function for the $q^4$ configuration is worked out as
\indent
\begin{eqnarray}\psi = \frac{1}{\sqrt{3}}
\left( \psi^{c}_{[211]_\lambda} \psi^{osf}_{[31]_\rho} -
\psi^{c}_{[211]_\rho} \psi^{osf}_{[31]_\lambda} +
\psi^{c}_{[211]_\eta} \psi^{osf}_{[31]_\eta} \right)
\end{eqnarray}
The detailed configurations of the spatial-spin-flavor as well as spin-flavor wave functions are worked out in the form of a Yamanouchi basis, as shown in Appendix~\ref{sec:AP1}.
The explicit forms of the color, spin, and flavor wave functions of pentaquarks are specified in the previous studies (see Refs. \cite{Sorakrai2012,Kai2014}).

\subsection{Spatial wave function}
We construct the spatial wave functions of the $q^4\bar q$ pentaquark systems in the harmonic oscillator potential for the quark-quark interaction.
The relative Jacobi coordinates and the corresponding momenta may be defined respectively as
\begin{eqnarray}
\vec{x}_{i}&=&\frac{i}{\sqrt{i+i^2}}(\frac{\sum_{j=1}^{i}m_j \vec{r}_j}{m_1+m_2+...m_i}-\vec{r}_{i +1}), \nonumber\\
\vec{p}_{i}&=& u_i \frac{d\vec x_i}{dt},
\end{eqnarray}
where $u_i$ are the reduced quark masses defined as
\begin{eqnarray}
u_i &=&\frac{(i+1)(\sum_{j=1}^{i} m_j)m_{i+1}}{i \sum_{j=1}^{i+1}m_j}, i=1,2,3,4
\end{eqnarray}
where $\vec{r}_{j}$ and $m_j$ are the coordinate and mass of the jth quark. We assign $x_1$, $x_2$, $x_3$, and $x_4$ to be $\rho$, $\lambda$, $\eta$, and $\xi$ Jacobi coordinates, respectively.

We start from the $q^4$ cluster. The $q^4$ spatial wave function, coupling among the $\rho$, $\lambda$, and $\eta$ harmonic oscillator wave functions,
may take the general form,
\begin{eqnarray}\label{eqn::spatial}
\psi^{q^4[X]_y}_{N'L'M'} &=&
\sum_{\{n_i,l_i\}}
 A(n_{\rho},n_{\lambda},n_{\eta},l_{\rho},l_{\lambda},l_{\eta}) \nonumber \\
&& \times \psi_{n_{\rho}l_{\rho}}(\vec\rho\,)\otimes\psi_{n_{\lambda}l_{\lambda}}(\vec\lambda\,) \otimes\psi_{n_{\eta}l_{\eta}}(\vec\eta\,)
\nonumber \\
&=& \sum_{\{n_i,l_i,m_i\}} C_{n_{\rho}, l_{\rho},m_{\rho},n_{\lambda},l_{\lambda},m_{\lambda},n_{\eta},l_{\eta},m_{\eta}} \nonumber \\
&& \times \psi_{n_{\rho}l_{\rho}m_{\rho}}(\vec\rho\,)\psi_{n_{\lambda}l_{\lambda}m_{\lambda}}(\vec\lambda\,)\psi_{n_{\eta}l_{\eta}m_{\eta}}(\vec\eta\,)
\end{eqnarray}
where $\psi_{n_{i}l_{i}m_{i}}$ are just harmonic oscillator wave functions and the sum $\{n_i,l_i\}$ is over $n_{\rho},n_{\lambda},n_{\eta}, l_{\rho},l_{\lambda},l_{\eta}$. $N'$, $L'$, and $M'$ are respectively the total principle quantum number, total angular momentum, and magnetic quantum number of the $q^4$ cluster. One has $N'= (2n_{\rho}+ l_{\rho})+(2n_{\lambda}+l_{\lambda})+(2n_{\eta}+ l_{\eta})$. The $[X]$ and $y$ in the superscript $[X]_y$ represent the irreducible representation $[X]$ and the $y$-type symmetry of the representation.
The coupling coefficients $A(n_{\rho},n_{\lambda},n_{\eta},l_{\rho},l_{\lambda},l_{\eta})$ as well as $C_{n_{\rho}, l_{\rho},m_{\rho},n_{\lambda},l_{\lambda},m_{\lambda},n_{\eta},l_{\eta},m_{\eta}}$ shall be determined according
to the $[X]_y$, where the representation matrices of the permutations of the $S_4$ group are applied to both sides of the general form.
The explicit forms of the spatial wave functions for the $q^4$ cluster are presented in Appendix~\ref{sec:AP2} for the permutation symmetries $\{{[4]_S}\}$. To save space, the other possible permutation symmetries $\{{[31]_{\rho,\lambda,\eta}}$, ${[211]_{\rho,\lambda,\eta}}$ and ${[22]_{\rho,\lambda}}\}$ will not be specified here.

The spatial wave function of pentaquark states is simply the product of the $q^4$ wave function and the harmonic oscillator wave function for the fourth Jacobi coordinate $\xi$,
where the antiquark is assigned the coordinate $\vec r_5$. The permutation symmetry of pentaquarks is simply represented by the $q^4$ cluster since $\psi_{n_\xi,l_\xi}(\vec\xi)$ is fully symmetric for any permutation between quarks. The total spatial wave function of pentaquarks may take the form,
\begin{eqnarray}\label{eqn::swf}
 \Psi_{NLM}^{[X]_y}= \psi^{q^4[X]_y}_{N'L'M'}\otimes\psi_{n_\xi,l_\xi}(\vec\xi)
\end{eqnarray}
where $\psi_{n_\xi,l_\xi}(\vec\xi)$ are the harmonic oscillator functions for the Jacobi coordinate $\xi$ and
$[X]_{y}$ stand for all possible permutation symmetries of the $q^4$ cluster, that is, $[X]_{y} = \{{[4]_S},\;{[31]_{\rho,\lambda,\eta}},\;{[211]_{\rho,\lambda,\eta}},\;{[22]_{\rho,\lambda}}\}$.
$N$, $L$, and $M$ are respectively the total principle quantum number, total angular momentum and magnetic quantum number of the pentaquark, with
\begin{eqnarray}\label{eqn::q4xi}
N= 2n_{\rho}+ l_{\rho}+2n_{\lambda}+l_{\lambda}+2n_{\eta}+ l_{\eta}+2n_{\xi}+l_{\xi}
\end{eqnarray}

In principle, one can construct the spatial wave functions of pentaquarks to any order by applying the representation matrices of the permutations of the $S_4$ group to the general form
in Eq. (\ref{eqn::spatial}). Though we have been dealing with a system where the quark-quark potential is the harmonic oscillator interaction, the spatial wave functions grouped in this work
according to the permutation symmetry can be employed as complete bases to study a system with other interactions.

\section{NUMERICAL CALCULATIONS OF THE PENTAQUARK MASS SPECTRUM}\label{sec:NC}

\indent We apply, as complete bases, the full wave functions of pentaquarks worked out in the previous section to study the pentaquark system described by the Hamiltonian
\begin{flalign}\label{eqn::ham}
&H =H_0+ H_{hyp}^{OGE}, \nonumber \\
&H_{0} =\sum_{k=1}^{N} (m_k+\frac{p_k^2}{2m_{k}})+\sum_{i<j}^{N}(-\frac{3}{8}\lambda^{C}_{i}\cdot\lambda^{C}_{j})(A_{ij} r_{ij}-\frac{B_{ij}}{r_{ij}}),  \nonumber \\
&H_{hyp}^{OGE} = -C_{OGE}\sum_{i<j}\frac{\lambda^{C}_{i}\cdot\lambda^{C}_{j}}{m_{i}m_{j}}\,\vec\sigma_{i}\cdot\vec\sigma_{j},
\end{flalign}
where $A_{ij}$ and $B_{ij}$ are mass-dependent coupling parameters, taking the form,
\begin{eqnarray}
A_{ij}= a \sqrt{\frac{m_{ij}}{m_u}},\;\;B_{ij}=b \sqrt{\frac{m_u}{m_{ij}}}
\end{eqnarray} 
with $m_{ij}$ being the reduced mass of ith and jth quarks, defined as $m_{ij}=\frac{2 m_i m_j}{m_i+m_j}$.
The hyperfine interaction, $H_{hyp}^{OGE}$, includes only one-gluon exchange contribution,
where $C_{OGE} = C_m\,m_u^2$, with $m_u$ being the constituent $u$ quark mass and $C_m$ a constant. $\lambda^C_{i}$ in the above equations are the generators of the color SU(3) group.

\indent The model parameters are determined by fitting the theoretical results to the experimental data of the masses
of the eight baryon isospin states, namely, $N(938),\; \Lambda(1115),\; \Sigma(1190),\; \Xi(1320)$, $\Delta(1232)$, $\Sigma^{*}(1385)$, $\Xi^{*}(1530)$, and $\Omega^{-}(1672)$,
as well as the first radial excitation state $N(1440)$ (Roper resonance) and a number of orbital excitation $l=1$ and $l=2$ baryons.
All these baryons are believed to be mainly $3q$ states whose masses were taken from the Particle Data Group \cite{PDG}.
The model parameters are predetermined as
\begin{eqnarray}\label{eq:nmo}
& 
m_u = m_d = 350 \, {\rm MeV},\,m_s = 525 \, {\rm MeV}, \nonumber\\
&
C_m   =  18 \, {\rm MeV},\, a     = 42000 \, {\rm MeV^2},\, b  =  0.72
\end{eqnarray}

\indent The Schr\"odinger equation for the pentaquark systems described by the Hamiltonian in Eq. (\ref{eqn::ham}) is solved numerically by expanding the pentaquark wave functions in the completed bases presented in Appendices~\ref{sec:AP2} and~\ref{sec:AP3}. For the ground state pentaquarks, one may not expect any orbital excitation, and hence the spatial wave functions should be fully symmetric.
Therefore, the spatial wave functions of pentaquarks states may be expanded in the $[4]_S$ complete basis:
\begin{eqnarray}\label{eqn::swf}
\Psi^{q^{4}\overline q}_{[4]_S}&=& a_1\psi^{q^4}_{000_{[4]_S}}\psi_{0,0}(\vec\xi)+a_2\psi^{q^4}_{200_{[4]_S}}\psi_{0,0}(\vec\xi) \nonumber \\
&&+\,a_3\psi^{q^4}_{000_{[4]_S}}\psi_{1,0}(\vec\xi) +a_4\psi^{q^4}_{400_{[4]_S}}\psi_{0,0}(\vec\xi)
 \nonumber \\
&&+\,a_5\psi^{q^4}_{200_{[4]_S}}\psi_{1,0}(\vec\xi)+a_6\psi^{q^4}_{000_{[4]_S}}\psi_{2,0}(\vec\xi)
  \nonumber \\
&&+\,a_7\psi^{q^4}_{600_{[4]_S}}\psi_{0,0}(\vec\xi)+a_8\psi^{q^4}_{400_{[4]_S}}\psi_{1,0}(\vec\xi)
 \nonumber \\
&&\vdots \nonumber \\
&&+\,a_{49}\psi^{q^4}_{1600_{[4]_S}}\psi_{3,0}(\vec\xi)+a_{50}\psi^{q^4}_{1400_{[4]_S}}\psi_{4,0}(\vec\xi) \nonumber \\
\end{eqnarray}
The mass spectra of the ground state $q^4\bar q$ and $q^3s\bar s$ pentaquarks are presented in Tables \ref{tab:d1} and \ref{tab:d2}.


\begin{table}[htb]
\caption{$q^4\bar q$ ground state pentaquark masses.}
\begin{center}\label{tab:d1}
\begin{tabular}{lcccccc}
 \hline
 \hline
  \\
& $q^4\bar q$ configurations && $J^{P}$ &&  $M(q^{4}\overline q)$  ({\rm MeV})
\\
\hline
 \\
& $\Psi^{csf}_{[211]_{C}[31]_{FS}[4]_{F}[31]_{S}}(q^{4}\overline q)$ && $\frac{1}{2}^{-}$, $\frac{3}{2}^{-}$
 && 2537, 2249
\\& \\
& $\Psi^{csf}_{[211]_{C}[31]_{FS}[31]_{F}[4]_{S}}(q^{4}\overline q)$ && $\frac{3}{2}^{-}$, $\frac{5}{2}^{-}$
 && 2009, 2249
\\&\\
& $\Psi^{csf}_{[211]_{C}[31]_{FS}[31]_{F}[31]_{S}}(q^{4}\overline q)$ && $\frac{1}{2}^{-}$,
$\frac{3}{2}^{-}$  && 2105, 2033
\\&\\
& $\Psi^{csf}_{[211]_{C}[31]_{FS}[31]_{F}[22]_{S}}(q^{4}\overline q)$ && $\frac{1}{2}^{-}$   && 2009
\\& \\
& $\Psi^{csf}_{[211]_{C}[31]_{FS}[22]_{F}[31]_{S}}(q^{4}\overline q)$ && $\frac{1}{2}^{-}$,
$\frac{3}{2}^{-}$  &&  1673, 2033 \\
\hline
\hline
\end{tabular}
\end{center}
\end{table}

\begin{table}[!t]
\caption{$q^3 s\bar s$ ground state pentaquark masses.}
\vspace*{-0.2cm}
\begin{center}\label{tab:d2}
\begin{tabular}{lcccccc}
\hline
\hline
\\
& $q^4\bar q$ configuration && $J^{P}$ &&  $M(q^{4}\overline q)$({\rm MeV})
\\
\hline
 \\
& $\Psi^{csf}_{[211]_{C}[31]_{FS}[4]_{F}[31]_{S}}(q^3s\bar s)$ && $\frac{1}{2}^{-}$, $\frac{3}{2}^{-}$
&&  2756, 2580
\\&\\
& $\Psi^{csf}_{[211]_{C}[31]_{FS}[31]_{F}[4]_{S}}(q^3s\bar s)$ && $\frac{3}{2}^{-}$, $\frac{5}{2}^{-}$
 && 2415, 2542
\\&\\
& $\Psi^{csf}_{[211]_{C}[31]_{FS}[31]_{F}[31]_{S}}(q^3s\bar s)$ && $\frac{1}{2}^{-}$,
$\frac{3}{2}^{-}$  && 2444, 2410
\\&\\
& $\Psi^{csf}_{[211]_{C}[31]_{FS}[31]_{F}[22]_{S}}(q^3s\bar s)$ && $\frac{1}{2}^{-}$   && 2389
\\&\\
& $\Psi^{csf}_{[211]_{C}[31]_{FS}[211]_{F}[31]_{S}}(q^3s\bar s)$ && $\frac{1}{2}^{-}$,
$\frac{3}{2}^{-}$   && 2030, 2242
\\&\\
& $\Psi^{csf}_{[211]_{C}[31]_{FS}[211]_{F}[22]_{S}}(q^3s\bar s)$ && $\frac{1}{2}^{-}$  &&
2164
\\&\\
& $\Psi^{csf}_{[211]_{C}[31]_{FS}[22]_{F}[31]_{S}}(q^3s\bar s)$ && $\frac{1}{2}^{-}$,
$\frac{3}{2}^{-}$   && 2132, 2352\\
\hline
\hline
\end{tabular}
\end{center}
\end{table}

It is noted that the state with the ${[31]_{FS}[22]_{F}[31]_{S}}$ configuration and quantum numbers $I(J^P)=\frac1{2}(\frac1{2}^{-})$ is predicted to
be the lowest pentaquark state. All nucleon resonances below $1.8 {\;\rm GeV}$ are more or less clear in experiment except for an isospin-$1/2$ narrow resonance, $N^{+}(1685)$, which was first announced in the photoproduction of an $\eta$ meson off a quasi free neutron \cite{Kuzn2007} and confirmed in 2017 \cite{Kuzn2017}, and also observed in the $\gamma d \rightarrow \eta n(p)$ excitation by two other groups \cite{Jaeg2011,Akon2014}. One may expect that the ${[31]_{FS}[22]_{F}[31]_{S}}$ pentaquark predicted in the work may be related to the $N(1685)$ since it is excluded from the normal nucleon resonances \cite{Sen2016,Ani2017}. 

In the wave of the observation of $\Theta^{+}(1540)$, a number of studies have tried to interpret the resonance as a $uudd\bar s$ state and explain its mass at $1.4-1.6$ ${\rm GeV}$ \cite{Kar2003,Jaffe2003,Ellis2004,Zhu2003,Sasaki2004}. Therefore, those works have predicted lower pentaquark masses than this work. 
In this work a potential model based on the Cornell-like potential shown in Eq. (\ref{eqn::ham}) is employed to estimate the pentaquark mass spectra with the model parameters predetermined by comparing the theoretical and experimental masses of low-lying baryons which are believed to be mainly $3q$ states. Except for the quark masses, three model parameters including the hyperfine interaction coupling constant are applied in this model to derive the baryon and pentaquark mass spectrums with the complete pentaquark wave functions. Comparing to other works \cite{Bijker2004,Bijker2019,Huang2018} for $q^4 \bar q$, hidden charm, and hidden strange pentaquark states separately, this model employs many fewer model parameters.

%
%
%
%
%
%

\section{Summary}\label{sec:SUM}
\indent In the work we have worked out all the spatial-spin-flavor as well as spin-flavor configurations of pentaquark systems, and derived explicitly
the spatial-spin-flavor as well as spin-flavor wave functions. Spatial wave functions of various permutation symmetries have been constructed for the pentaquark systems where the quark-quark interactions are of the harmonic oscillator type. The constructed spatial wave functions can serve as complete bases for studying systems in other interactions.

As a simple application, the $q^4\bar q$ and $q^3 s\bar s$ pentaquarks are studied in more realistic potentials, and the masses of the ground state pentaquarks are accurately evaluated. It is found that the ground state pentaquark masses are predicted, in the Cornell-like potential in this work, to be relatively higher than the masses from other theoretical works. However, it is interesting to 
note that the work predicts that the pentaquark state with the ${[31]_{FS}[22]_{F}[31]_{S}}$ configuration and the quantum numbers $I(J^P)=\frac1{2}(\frac1{2}^{-})$ 
has the lowest mass, about 1670 MeV, which is quite close the mass of the isospin-$1/2$ narrow resonance $N^{+}(1685)$. 

\section{ACKNOWLEDGMENTS}
\indent  This work is supported by Suranaree University of Technology (SUT) and the Office of the Higher Education Commission under the NRU project of Thailand. K.X. and Y.Y. acknowledge support from SUT under Grant No. SUT-PhD/13/2554. A.K. and  A.L. acknowledge support from SUT. S.S. acknowledges support from the Faculty of Science, Burapha University. X.Y.L. acknowledges support from Young Science Foundation from the Education Department of Liaoning Province, China (Project No. LQ2019009).

\begin{widetext}

\appendix
\section{Explicit spatial-spin-flavor wave function of the pentaquark}\label{sec:AP1}
In this appendix we list explicitly the spatial-spin-flavor as well as spin-flavor wave functions for all the configurations in Tables \ref{tab:wf0osf} and \ref{tab:wf0sf}.
$O_{[X]}$, $F_{[X]}$, $S_{[X]}$, $FS_{[X]}$, and $OFS_{[X]}$ in the content below stand respectively for the spatial, flavor, spin, spin-flavor, and spatial-spin-flavor symmetries.  $OFS_{[31]}$ and $FS_{[4]}$, $FS_{[31]}$, $FS_{[211]}$, $FS_{[22]}$, ${FS}_{[1111]}$ wave functions are listed separately in Tables \ref{con1}, \ref{con2}, \ref{con3}, \ref{con4}, \ref{con5} and \ref{con6}.

\begin{longtable}{|l|l|l|}
\caption[Explicit $OFS_{[31]}$ wave function]{Explicit $OFS_{[31]}$ wave functions.} \label{con1} \\

\hline \multicolumn{1}{|c|}{\textbf{$O_{[X]}$ $FS_{[X]}$}} & \multicolumn{1}{|c|}{$OFS_{[31]}$ Type}& \multicolumn{1}{c|}{Explicit wave function}  \\ \hline
\endfirsthead

\multicolumn{3}{c}%
{{\bfseries \tablename\ \thetable{} -- continued from previous page}} \\
\hline \multicolumn{1}{|c|}{\textbf{$O_{[X]}$ $FS_{[X]}$}} & \multicolumn{1}{|c|}{$OFS_{[31]}$ Type}
& \multicolumn{1}{c|}{Explicit wave function} \\ \hline
\endhead

\hline \multicolumn{3}{|r|}{{Continued on next page}} \\ \hline
\endfoot

\hline \hline
\endlastfoot

$O_{[4]} FS_{[31]}$ & $\rho$ & $ \phi^o_{[4]_S}
\phi^{sf}_{[31]_{\rho}}$ \\
\phantom{} & $\lambda$ & $ \phi^o_{[4]_S}
\phi^{sf}_{[31]_{\lambda}}$  \\
\phantom{} & $\eta$ & $\phi^o_{[4]_S}
\phi^{sf}_{[31]_{\eta}}$ \\
 \hline

$O_{[1111]} FS_{[211]}$ & $\rho$ & $\phi^o_{[1111]_A}
\phi^{sf}_{[211]_{\rho}}$ \\
\phantom{} & $\lambda$ & $\phi^o_{[1111]_A}
\phi^{sf}_{[211]_{\lambda}}$  \\
\phantom{} & $\eta$ & $ \phi^o_{[1111]_A}
\phi^{sf}_{[211]_{\eta}}$ \\
 \hline


$O_{[22]} FS_{[31]}$ & $\rho$ & $\frac{1}{2}
\phi^o_{[22]_{\lambda}} \phi^{sf}_{[31]_{\rho}} +\frac{1}{2}
\phi^o_{[22]_{\rho}} \phi^{sf}_{[31]_{\lambda}} -\frac{1}{\sqrt{2}}
\phi^o_{[22]_{\rho}} \phi^{sf}_{[31]_{\eta}}$ \\
\phantom{} & $\lambda$ & $ -\frac{1}{2}
\phi^o_{[22]_{\lambda}} \phi^{sf}_{[31]_{\lambda}} +\frac{1}{2}
\phi^o_{[22]_{\rho}} \phi^{sf}_{[31]_{\rho}} -\frac{1}{\sqrt{2}}
\phi^o_{[22]_{\lambda}} \phi^{sf}_{[31]_{\eta}}$  \\
\phantom{} & $\eta$ & $-\frac{1}{\sqrt{2}} \phi^o_{[22]_{\lambda}} \phi^{sf}_{[31]_{\lambda}}
- \frac{1}{\sqrt{2}} \phi^o_{[22]_{\rho}} \phi^{sf}_{[31]_{\rho}}$ \\
 \hline

$O_{[22]} FS_{[211]}$ & $\rho$ & $ \frac{1}{2}
 \phi^o_{[22]_{\rho}} \phi^{sf}_{[211]_{\lambda}}+\frac{1}{2}
\phi^o_{[22]_{\lambda}} \phi^{sf}_{[211]_{\rho}} +\frac{1}{\sqrt{2}}
 \phi^o_{[22]_{\lambda}} \phi^{sf}_{[211]_{\eta}}$ \\
\phantom{} & $\lambda$ & $-\frac{1}{2}
 \phi^o_{[22]_{\lambda}} \phi^{sf}_{[211]_{\lambda}}+\frac{1}{2}
 \phi^o_{[22]_{\rho}}\phi^{sf}_{[211]_{\rho}} -\frac{1}{\sqrt{2}}
 \phi^o_{[22]_{\rho}}\phi^{sf}_{[211]_{\eta}}$  \\
\phantom{} & $\eta$ & $\frac{1}{\sqrt{2}}  \phi^o_{[22]_{\lambda}}\phi^{sf}_{[211]_{\lambda}}+
 \frac{1}{\sqrt{2}}  \phi^o_{[22]_{\rho}}\phi^{sf}_{[211]_{\rho}}$ \\
 \hline

$O_{[211]} FS_{[31]}$ & $\rho$ & $\frac{1}{\sqrt{2}} \phi^o_{[211]_{\eta}} \phi^{sf}_{[31]_{\lambda}}
+ \frac{1}{\sqrt{2}} \phi^o_{[211]_{\rho}} \phi^{sf}_{[31]_{\eta}}$ \\
\phantom{} & $\lambda$ & $\frac{1}{\sqrt{2}}
\phi^o_{[211]_{\lambda}} \phi^{sf}_{[31]_{\eta}} -\frac{1}{\sqrt{2}}
\phi^o_{[211]_{\eta}} \phi^{sf}_{[31]_{\rho}}$  \\
\phantom{} & $\eta$ & $-\frac{1}{\sqrt{2}} \phi^o_{[211]_{\lambda}} \phi^{sf}_{[31]_{\lambda}}
- \frac{1}{\sqrt{2}} \phi^o_{[211]_{\rho}} \phi^{sf}_{[31]_{\rho}}$ \\
 \hline

$O_{[211]} FS_{[211]}$ & $\rho$ & $ \frac{1}{\sqrt{3}}
\phi^o_{[211]_{\lambda}} \phi^{sf}_{[211]_{\rho}} +\frac{1}{\sqrt{3}}
\phi^o_{[211]_{\rho}} \phi^{sf}_{[211]_{\lambda}} -\frac{1}{\sqrt{6}}
\phi^o_{[211]_{\eta}} \phi^{sf}_{[211]_{\lambda}}-\frac{1}{\sqrt{6}}
\phi^o_{[211]_{\lambda}} \phi^{sf}_{[211]_{\eta}}$ \\
\phantom{} & $\lambda$ & $ -\frac{1}{\sqrt{3}}
\phi^o_{[211]_{\lambda}} \phi^{sf}_{[211]_{\lambda}} +\frac{1}{\sqrt{3}}
\phi^o_{[211]_{\rho}} \phi^{sf}_{[211]_{\rho}} +\frac{1}{\sqrt{6}}
\phi^o_{[211]_{\eta}} \phi^{sf}_{[211]_{\rho}}+\frac{1}{\sqrt{6}}
\phi^o_{[211]_{\rho}} \phi^{sf}_{[211]_{\eta}}$  \\
\phantom{} & $\eta$ & $-\frac{1}{\sqrt{6}} \phi^o_{[211]_{\lambda}} \phi^{sf}_{[211]_{\lambda}}
- \frac{1}{\sqrt{6}} \phi^o_{[211]_{\rho}} \phi^{sf}_{[211]_{\rho}}
+\frac{2}{\sqrt{6}} \phi^o_{[211]_{\eta}} \phi^{sf}_{[211]_{\eta}}$ \\
 \hline

 $O_{[211]} FS_{[22]}$ & $\rho$ & $ \frac{1}{2}
\phi^o_{[211]_{\lambda}}\phi^{sf}_{[22]_{\rho}} +\frac{1}{2}
\phi^o_{[211]_{\rho}} \phi^{sf}_{[22]_{\lambda}} +\frac{1}{\sqrt{2}}
\phi^o_{[211]_{\eta}} \phi^{sf}_{[22]_{\lambda}}$ \\
\phantom{} & $\lambda$ & $-\frac{1}{2}
\phi^o_{[211]_{\lambda}} \phi^{sf}_{[22]_{\lambda}} +\frac{1}{2}
\phi^o_{[211]_{\rho}} \phi^{sf}_{[22]_{\rho}} -\frac{1}{\sqrt{2}}
\phi^o_{[211]_{\eta}} \phi^{sf}_{[22]_{\rho}}$  \\
\phantom{} & $\eta$ & $\frac{1}{\sqrt{2}} \phi^o_{[211]_{\lambda}} \phi^{sf}_{[22]_{\lambda}}+
 \frac{1}{\sqrt{2}} \phi^o_{[211]_{\rho}} \phi^{sf}_{[22]_{\rho}}$ \\
 \hline

 $O_{[31]} FS_{[4]}$ & $\rho$ & $ \phi^o_{[31]_{\rho}}
\phi^{sf}_{[4]_S}$ \\
\phantom{} & $\lambda$ & $ \phi^o_{[31]_{\lambda}}
\phi^{sf}_{[4]_S}$  \\
\phantom{} & $\eta$ & $\phi^o_{[31]_{\eta}}
\phi^{sf}_{[4]_S}$ \\
 \hline

 $O_{[31]} FS_{[31]}$ & $\rho$ & $\frac{1}{\sqrt{3}}
\phi^o_{[31]_{\lambda}} \phi^{sf}_{[31]_{\rho}} +\frac{1}{\sqrt{3}}
\phi^o_{[31]_{\rho}} \phi^{sf}_{[31]_{\lambda}} +\frac{1}{\sqrt{6}}
\phi^o_{[31]_{\eta}} \phi^{sf}_{[31]_{\rho}}+\frac{1}{\sqrt{6}}
\phi^o_{[31]_{\rho}} \phi^{sf}_{[31]_{\eta}}$ \\
\phantom{} & $\lambda$ & $-\frac{1}{\sqrt{3}}
\phi^o_{[31]_{\lambda}} \phi^{sf}_{[31]_{\lambda}} +\frac{1}{\sqrt{3}}
\phi^o_{[31]_{\rho}} \phi^{sf}_{[31]_{\rho}} +\frac{1}{\sqrt{6}}
\phi^o_{[31]_{\eta}} \phi^{sf}_{[31]_{\lambda}}+\frac{1}{\sqrt{6}}
\phi^o_{[31]_{\lambda}} \phi^{sf}_{[31]_{\eta}}$  \\
\phantom{} & $\eta$ & $\frac{1}{\sqrt{6}} \phi_{[31]_{\lambda}} \phi^{sf}_{[31]_{\lambda}}
+ \frac{1}{\sqrt{6}} \phi_{[31]_{\rho}} \phi^{sf}_{[31]_{\rho}}
-\frac{2}{\sqrt{6}} \phi_{[31]_{\eta}} \phi^{sf}_{[31]_{\eta}}$ \\
 \hline

  $O_{[31]} FS_{[211]}$ & $\rho$ & $ \frac{1}{\sqrt{2}} \phi^o_{[31]_{\lambda}} \phi^{sf}_{[211]_{\eta}}
+ \frac{1}{\sqrt{2}}  \phi^o_{[31]_{\eta}}\phi^{sf}_{[211]_{\rho}}$ \\
\phantom{} & $\lambda$ & $ \frac{1}{\sqrt{2}}
 \phi^o_{[31]_{\eta}} \phi^{sf}_{[211]_{\lambda}}-\frac{1}{\sqrt{2}}
\phi^o_{[31]_{\rho}}\phi^{sf}_{[211]_{\eta}}$  \\
\phantom{} & $\eta$ & $-
\frac{1}{\sqrt{2}} \phi^o_{[31]_{\lambda}}\phi^{sf}_{[211]_{\lambda}}
- \frac{1}{\sqrt{2}}  \phi^o_{[31]_{\rho}}\phi^{sf}_{[211]_{\rho}}$ \\
 \hline

  $O_{[31]} FS_{[22]}$ & $\rho$ & $\frac{1}{2}
\phi^o_{[31]_{\lambda}} \phi^{sf}_{[22]_{\rho}} +\frac{1}{2}
\phi^o_{[31]_{\rho}} \phi^{sf}_{[22]_{\lambda}} -\frac{1}{\sqrt{2}}
\phi^o_{[31]_{\eta}} \phi^{sf}_{[22]_{\rho}}$ \\
\phantom{} & $\lambda$ & $-\frac{1}{2}
\phi^o_{[31]_{\lambda}} \phi^{sf}_{[22]_{\lambda}} +\frac{1}{2}
\phi^o_{[31]_{\rho}} \phi^{sf}_{[22]_{\rho}} -\frac{1}{\sqrt{2}}
\phi^o_{[31]_{\eta}} \phi^{sf}_{[22]_{\lambda}}$  \\
\phantom{} & $\eta$ & $-\frac{1}{\sqrt{2}} \phi^o_{[31]_{\lambda}} \phi^{sf}_{[22]_{\lambda}}
- \frac{1}{\sqrt{2}} \phi^o_{[31]_{\rho}} \phi^{sf}_{[22]_{\rho}}$ \\
\end{longtable}

\begin{longtable}{|l|l|l|}
\caption[Explicit $FS_{[4]}$ wave function]{Explicit $FS_{[4]}$ wave function} \label{con2} \\

\hline \multicolumn{1}{|c|}{\textbf{$F_{[X]}$ $S_{[X]}$}} & \multicolumn{1}{c|}{$FS_{[4]}$ Type}& \multicolumn{1}{c|}{Explicit wave function}  \\ \hline
\endfirsthead

\multicolumn{3}{c}%
{{\bfseries \tablename\ \thetable{} -- continued from previous page}} \\
\hline \multicolumn{1}{|c|}{\textbf{$F_{[X]}$ $S_{[X]}$}} & \multicolumn{1}{c|}{$FS_{[4]}$ Type}
& \multicolumn{1}{c|}{Explicit wave function} \\ \hline
\endhead

\hline \multicolumn{3}{|r|}{{Continued on next page}} \\ \hline
\endfoot

\hline \hline
\endlastfoot

$F_{[4]} S_{[4]}$ & $S$ & $\phi_{[4]_S}
\chi_{[4]_S}$ \\
 \hline

$F_{[31]} S_{[31]}$ & $S$ & $\frac{1}{\sqrt{3}}\phi_{[31]_\lambda}
\chi_{[31]_\lambda}+\frac{1}{\sqrt{3}}\phi_{[31]_{\rho}}
\chi_{[31]_{\rho}} +\frac{1}{\sqrt{3}}\phi_{[31]_{\eta}} \chi_{[31]_{\eta}}$ \\
 \hline

$F_{[22]} S_{[22]}$ & $S$ & $\frac{1}{\sqrt{2}}\phi_{[22]_\lambda}
\chi_{[22]_\lambda}+\frac{1}{\sqrt{2}}\phi_{[22]_{\rho}}
\chi_{[22]_{\rho}}$ \\
\end{longtable}

\begin{longtable}{|l|l|l|}
\caption[Explicit $FS_{[31]}$ wave function]{Explicit $FS_{[31]}$ wave function} \label{con3} \\

\hline \multicolumn{1}{|c|}{\textbf{$F_{[X]}$ $S_{[X]}$}} & \multicolumn{1}{c|}{$FS_{[31]}$ Type}& \multicolumn{1}{c|}{Explicit wave function}  \\ \hline
\endfirsthead

\multicolumn{3}{c}%
{{\bfseries \tablename\ \thetable{} -- continued from previous page}} \\
\hline \multicolumn{1}{|c|}{\textbf{$F_{[X]}$ $S_{[X]}$}} & \multicolumn{1}{c|}{$FS_{[31]}$ Type}
& \multicolumn{1}{c|}{Explicit wave function} \\ \hline
\endhead

\hline \multicolumn{3}{|r|}{{Continued on next page}} \\ \hline
\endfoot

\hline \hline
\endlastfoot

$F_{[4]} S_{[31]}$ & $\rho$ & $\phi_{[4]_S}
\chi_{[31]_{\rho}}$ \\
\phantom{} & $\lambda$ & $\phi_{[4]_S}
\chi_{[31]_{\lambda}}$  \\
\phantom{} & $\eta$ & $\phi_{[4]_S}
\chi_{[31]_{\eta}}$ \\
 \hline

$F_{[31]} S_{[4]}$ & $\rho$ & $\phi_{[31]_{\rho}}
\chi_{[4]_S}$ \\
\phantom{} & $\lambda$ & $\phi_{[31]_{\lambda}}
\chi_{[4]_S}$  \\
\phantom{} & $\eta$ & $\phi_{[31]_{\eta}}
\chi_{[4]_S}$ \\
 \hline

$F_{[31]} S_{[31]}$ & $\rho$ & $\frac{1}{\sqrt{3}}
\phi_{[31]_{\lambda}} \chi_{[31]_{\rho}} +\frac{1}{\sqrt{3}}
\phi_{[31]_{\rho}} \chi_{[31]_{\lambda}} +\frac{1}{\sqrt{6}}
\phi_{[31]_{\eta}} \chi_{[31]_{\rho}}+\frac{1}{\sqrt{6}}
\phi_{[31]_{\rho}} \chi_{[31]_{\eta}}$ \\
\phantom{} & $\lambda$ & $-\frac{1}{\sqrt{3}}
\phi_{[31]_{\lambda}} \chi_{[31]_{\lambda}} +\frac{1}{\sqrt{3}}
\phi_{[31]_{\rho}} \chi_{[31]_{\rho}} +\frac{1}{\sqrt{6}}
\phi_{[31]_{\eta}} \chi_{[31]_{\lambda}}+\frac{1}{\sqrt{6}}
\phi_{[31]_{\lambda}} \chi_{[31]_{\eta}}$  \\
\phantom{} & $\eta$ & $\frac{1}{\sqrt{6}} \phi_{[31]_{\lambda}} \chi_{[31]_{\lambda}}
+ \frac{1}{\sqrt{6}} \phi_{[31]_{\rho}} \chi_{[31]_{\rho}}
-\frac{2}{\sqrt{6}} \phi_{[31]_{\eta}} \chi_{[31]_{\eta}}$ \\
 \hline

$F_{[31]} FS_{[22]}$ & $\rho$ & $  \frac{1}{2}
\phi_{[31]_{\lambda}} \chi_{[22]_{\rho}} +\frac{1}{2}
\phi_{[31]_{\rho}} \chi_{[22]_{\lambda}} -\frac{1}{\sqrt{2}}
\phi_{[31]_{\eta}} \chi_{[22]_{\rho}}$ \\
\phantom{} & $\lambda$ & $ -\frac{1}{2}
\phi_{[31]_{\lambda}} \chi_{[22]_{\lambda}} +\frac{1}{2}
\phi_{[31]_{\rho}} \chi_{[22]_{\rho}} -\frac{1}{\sqrt{2}}
\phi_{[31]_{\eta}} \chi_{[22]_{\lambda}}$  \\
\phantom{} & $\eta$ & $-\frac{1}{\sqrt{2}} \phi_{[31]_{\lambda}} \chi_{[22]_{\lambda}}
- \frac{1}{\sqrt{2}} \phi_{[31]_{\rho}} \chi_{[22]_{\rho}}$ \\
 \hline

$F_{[211]} S_{[31]}$ & $\rho$ & $\frac{1}{\sqrt{2}} \phi_{[211]_{\eta}} \chi_{[31]_{\lambda}}
+ \frac{1}{\sqrt{2}} \phi_{[211]_{\rho}} \chi_{[31]_{\eta}}$ \\
\phantom{} & $\lambda$ & $\frac{1}{\sqrt{2}}
\phi_{[211]_{\lambda}} \chi_{[31]_{\eta}} -\frac{1}{\sqrt{2}}
\phi_{[211]_{\eta}} \chi_{[31]_{\rho}}$  \\
\phantom{} & $\eta$ & $-\frac{1}{\sqrt{2}} \phi_{[211]_{\lambda}} \chi_{[31]_{\lambda}}
- \frac{1}{\sqrt{2}} \phi_{[211]_{\rho}} \chi_{[31]_{\rho}}$ \\
 \hline

$F_{[211]} S_{[22]}$ & $\rho$ & $ \frac{1}{2}
\phi_{[211]_{\lambda}} \chi_{[22]_{\rho}} +\frac{1}{2}
\phi_{[211]_{\rho}} \chi_{[22]_{\lambda}} +\frac{1}{\sqrt{2}}
\phi_{[211]_{\eta}} \chi_{[22]_{\lambda}}$ \\
\phantom{} & $\lambda$ & $ -\frac{1}{2}
\phi_{[211]_{\lambda}} \chi_{[22]_{\lambda}} +\frac{1}{2}
\phi_{[211]_{\rho}} \chi_{[22]_{\rho}} -\frac{1}{\sqrt{2}}
\phi_{[211]_{\eta}} \chi_{[22]_{\rho}}$  \\
\phantom{} & $\eta$ & $\frac{1}{\sqrt{2}} \phi_{[211]_{\lambda}} \chi_{[22]_{\lambda}}+
 \frac{1}{\sqrt{2}} \phi_{[211]_{\rho}} \chi_{[22]_{\rho}}$ \\
 \hline

 $F_{[22]} S_{[31]}$ & $\rho$ & $\frac{1}{2}
\phi_{[22]_{\lambda}} \chi_{[31]_{\rho}} +\frac{1}{2}
\phi_{[22]_{\rho}} \chi_{[31]_{\lambda}} -\frac{1}{\sqrt{2}}
\phi_{[22]_{\rho}} \chi_{[31]_{\eta}}$ \\
\phantom{} & $\lambda$ & $ -\frac{1}{2}
\phi_{[22]_{\lambda}} \chi_{[31]_{\lambda}} +\frac{1}{2}
\phi_{[22]_{\rho}} \chi_{[31]_{\rho}} -\frac{1}{\sqrt{2}}
\phi_{[22]_{\lambda}} \chi_{[31]_{\eta}}$  \\
\phantom{} & $\eta$ & $-
\frac{1}{\sqrt{2}} \phi_{[22]_{\lambda}} \chi_{[31]_{\lambda}}
- \frac{1}{\sqrt{2}} \phi_{[22]_{\rho}} \chi_{[31]_{\rho}}$ \\
 \end{longtable}

\begin{longtable}{|l|l|l|}
\caption[Explicit $FS_{[211]}$ wave function]{Explicit $FS_{[211]}$ wave function} \label{con4} \\

\hline \multicolumn{1}{|c|}{\textbf{$F_{[X]}$ $S_{[X]}$}} & \multicolumn{1}{c|}{$FS_{[211]}$ Type}& \multicolumn{1}{c|}{Explicit wave function}  \\ \hline
\endfirsthead

\multicolumn{3}{c}%
{{\bfseries \tablename\ \thetable{} -- continued from previous page}} \\
\hline \multicolumn{1}{|c|}{\textbf{$F_{[X]}$ $S_{[X]}$}} & \multicolumn{1}{c|}{$FS_{[211]}$ Type}
& \multicolumn{1}{c|}{Explicit wave function} \\ \hline
\endhead

\hline \multicolumn{3}{|r|}{{Continued on next page}} \\ \hline
\endfoot

\hline \hline
\endlastfoot

$F_{[31]} S_{[31]}$ & $\rho$ & $\frac{1}{\sqrt{2}} \phi_{[31]_{\eta}} \chi_{[31]_{\rho}}
- \frac{1}{\sqrt{2}} \phi_{[31]_{\rho}} \chi_{[31]_{\eta}}$ \\
\phantom{} & $\lambda$ & $ \frac{1}{\sqrt{2}}
\phi_{[31]_{\eta}} \chi_{[31]_{\lambda}} -\frac{1}{\sqrt{2}}
\phi_{[31]_{\lambda}} \chi_{[31]_{\eta}}$  \\
\phantom{} & $\eta$ & $\frac{1}{\sqrt{2}} \phi_{[31]_{\lambda}} \chi_{[31]_{\rho}}
- \frac{1}{\sqrt{2}} \phi_{[31]_{\rho}} \chi_{[31]_{\lambda}}$ \\
 \hline

$F_{[31]} S_{[22]}$ & $\rho$ & $ \frac{1}{2}
\phi_{[31]_{\lambda}} \chi_{[22]_{\rho}} +\frac{1}{2}
\phi_{[31]_{\rho}} \chi_{[22]_{\lambda}} +\frac{1}{\sqrt{2}}
\phi_{[31]_{\eta}} \chi_{[22]_{\rho}}$ \\
\phantom{} & $\lambda$ & $ -\frac{1}{2}
\phi_{[31]_{\lambda}} \chi_{[22]_{\lambda}} +\frac{1}{2}
\phi_{[31]_{\rho}} \chi_{[22]_{\rho}} +\frac{1}{\sqrt{2}}
\phi_{[31]_{\eta}} \chi_{[22]_{\lambda}}$  \\
\phantom{} & $\eta$ & $\frac{1}{\sqrt{2}} \phi_{[31]_{\rho}} \chi_{[22]_{\lambda}}
- \frac{1}{\sqrt{2}} \phi_{[31]_{\lambda}} \chi_{[22]_{\rho}}$ \\
 \hline

$F_{[211]} S_{[4]}$ & $\rho$ & $ \phi_{[211]_{\rho}}
\chi_{[4]_S}$ \\
\phantom{} & $\lambda$ & $ \phi_{[211]_{\lambda}}
\chi_{[4]_S}$  \\
\phantom{} & $\eta$ & $ \phi_{[211]_{\eta}}
\chi_{[4]_S}$ \\
 \hline

$F_{[211]} S_{[31]}$ & $\rho$ & $\frac{1}{\sqrt{3}}
\phi_{[211]_{\lambda}} \chi_{[31]_{\rho}} +\frac{1}{\sqrt{3}}
\phi_{[211]_{\rho}} \chi_{[31]_{\lambda}} +\frac{1}{\sqrt{6}}
\phi_{[211]_{\eta}} \chi_{[31]_{\lambda}}-\frac{1}{\sqrt{6}}
\phi_{[211]_{\rho}} \chi_{[31]_{\eta}}$ \\
\phantom{} & $\lambda$ & $-\frac{1}{\sqrt{3}}
\phi_{[211]_{\lambda}} \chi_{[31]_{\lambda}} +\frac{1}{\sqrt{3}}
\phi_{[211]_{\rho}} \chi_{[31]_{\rho}} -\frac{1}{\sqrt{6}}
\phi_{[211]_{\eta}} \chi_{[31]_{\rho}}-\frac{1}{\sqrt{6}}
\phi_{[211]_{\lambda}} \chi_{[31]_{\eta}}$  \\
\phantom{} & $\eta$ & $\frac{1}{\sqrt{6}} \phi_{[211]_{\lambda}} \chi_{[31]_{\rho}}
- \frac{1}{\sqrt{6}} \phi_{[211]_{\rho}} \chi_{[31]_{\lambda}}
+\frac{2}{\sqrt{6}} \phi_{[211]_{\eta}} \chi_{[31]_{\eta}}$ \\
 \hline

$F_{[211]} FS_{[22]}$ & $\rho$ & $\frac{1}{2}
\phi_{[211]_{\lambda}} \chi_{[22]_{\rho}} +\frac{1}{2}
\phi_{[211]_{\rho}} \chi_{[22]_{\lambda}} -\frac{1}{\sqrt{2}}
\phi_{[211]_{\eta}} \chi_{[22]_{\lambda}}$ \\
\phantom{} & $\lambda$ & $-\frac{1}{2}
\phi_{[211]_{\lambda}} \chi_{[22]_{\lambda}} +\frac{1}{2}
\phi_{[211]_{\rho}} \chi_{[22]_{\rho}} +\frac{1}{\sqrt{2}}
\phi_{[211]_{\eta}} \chi_{[22]_{\rho}}$  \\
\phantom{} & $\eta$ & $-\frac{1}{\sqrt{2}} \phi_{[211]_{\rho}} \chi_{[22]_{\lambda}}+
 \frac{1}{\sqrt{2}} \phi_{[211]_{\lambda}} \chi_{[22]_{\rho}}$ \\
 \hline

$F_{[22]} S_{[31]}$ & $\rho$ & $\frac{1}{2}
\phi_{[22]_{\lambda}} \chi_{[31]_{\rho}} +\frac{1}{2}
\phi_{[22]_{\rho}} \chi_{[31]_{\lambda}} +\frac{1}{\sqrt{2}}
\phi_{[22]_{\rho}} \chi_{[31]_{\eta}}$ \\
\phantom{} & $\lambda$ & $ -\frac{1}{2}
\phi_{[22]_{\lambda}} \chi_{[31]_{\lambda}} +\frac{1}{2}
\phi_{[22]_{\rho}} \chi_{[31]_{\rho}} +\frac{1}{\sqrt{2}}
\phi_{[22]_{\lambda}} \chi_{[31]_{\eta}}$  \\
\phantom{} & $\eta$ & $-\frac{1}{\sqrt{2}} \phi_{[22]_{\rho}} \chi_{[31]_{\lambda}}
+ \frac{1}{\sqrt{2}} \phi_{[22]_{\lambda}} \chi_{[31]_{\rho}}$ \\
 \end{longtable}

\begin{longtable}{|l|l|l|}
\caption[Explicit $FS_{[22]}$ wave function]{Explicit $FS_{[22]}$ wave function} \label{con5} \\

\hline \multicolumn{1}{|c|}{$F_{[X]}$ $S_{[X]}$} & \multicolumn{1}{c|}{$FS_{[22]}$ Type}& \multicolumn{1}{c|}{Explicit wave function}  \\ \hline
\endfirsthead

\multicolumn{3}{c}%
{{ \tablename\ \thetable{} -- continued from previous page}} \\
\hline \multicolumn{1}{|c|}{$F_{[X]}$ $S_{[X]}$} & \multicolumn{1}{c|}{$FS_{[22]}$ Type}
& \multicolumn{1}{c|}{Explicit wave function} \\ \hline
\endhead

\hline \multicolumn{3}{|r|}{{Continued on next page}} \\ \hline
\endfoot

\hline \hline
\endlastfoot

$F_{[4]} S_{[22]}$ & $\rho$ & $\phi_{[4]_S}
\chi_{[22]_{\rho}}$ \\
\phantom{} & $\lambda$ & $\phi_{[4]_S}
\chi_{[22]_{\lambda}}$  \\
 \hline

$F_{[31]} S_{[31]}$ & $\rho$ & $\frac{1}{\sqrt{3}}
\phi_{[31]_{\lambda}} \chi_{[31]_{\rho}} +\frac{1}{\sqrt{3}}
\phi_{[31]_{\rho}} \chi_{[31]_{\lambda}} -\frac{1}{\sqrt{6}}
\phi_{[31]_{\eta}} \chi_{[31]_{\rho}}-\frac{1}{\sqrt{6}}
\phi_{[31]_{\rho}} \chi_{[31]_{\eta}}$ \\
\phantom{} & $\lambda$ & $-\frac{1}{\sqrt{3}}
\phi_{[31]_{\lambda}} \chi_{[31]_{\lambda}} +\frac{1}{\sqrt{3}}
\phi_{[31]_{\rho}} \chi_{[31]_{\rho}} -\frac{1}{\sqrt{6}}
\phi_{[31]_{\eta}} \chi_{[31]_{\lambda}}-\frac{1}{\sqrt{6}}
\phi_{[31]_{\lambda}} \chi_{[31]_{\eta}}$  \\
 \hline

$F_{[211]} S_{[31]}$ & $\rho$ & $\frac{1}{\sqrt{3}}
\phi_{[211]_{\lambda}} \chi_{[31]_{\rho}} +\frac{1}{\sqrt{3}}
\phi_{[211]_{\rho}} \chi_{[31]_{\lambda}} -\frac{1}{\sqrt{6}}
\phi_{[211]_{\eta}} \chi_{[31]_{\lambda}}+\frac{1}{\sqrt{6}}
\phi_{[211]_{\rho}} \chi_{[31]_{\eta}}$ \\
\phantom{} & $\lambda$ & $-\frac{1}{\sqrt{3}}
\phi_{[211]_{\lambda}} \chi_{[31]_{\lambda}} +\frac{1}{\sqrt{3}}
\phi_{[211]_{\rho}} \chi_{[31]_{\rho}} +\frac{1}{\sqrt{6}}
\phi_{[211]_{\eta}} \chi_{[31]_{\rho}}+\frac{1}{\sqrt{6}}
\phi_{[211]_{\lambda}} \chi_{[31]_{\eta}}$  \\
 \hline

$F_{[22]} S_{[4]}$ & $\rho$ & $\phi_{[22]_{\rho}}
\chi_{[4]_S}$ \\
\phantom{} & $\lambda$ & $\phi_{[22]_{\lambda}}
\chi_{[4]_S}$  \\
 \hline

$F_{[22]} S_{[22]}$ & $\rho$ & $\frac{1}{\sqrt{2}} \phi_{[22]_{\rho}} \chi_{[22]_{\lambda}}
+ \frac{1}{\sqrt{2}} \phi_{[22]_{\lambda}} \chi_{[22]_{\rho}}$ \\
\phantom{} & $\lambda$ & $-\frac{1}{\sqrt{2}}
\phi_{[22]_{\lambda}} \chi_{[22]_{\lambda}} +\frac{1}{\sqrt{2}}
\phi_{[22]_{\rho}} \chi_{[22]_{\rho}}$ \\
\end{longtable}

\begin{longtable}{|l|l|l|}
\caption[Explicit $FS_{[1111]}$ wave function]{Explicit $FS_{[1111]}$ wave function} \label{con6} \\

\hline \multicolumn{1}{|c|}{\textbf{$F_{[X]}$ $S_{[X]}$}} & \multicolumn{1}{c|}{$FS_{[1111]}$ Type}& \multicolumn{1}{c|}{Explicit wave function}  \\ \hline
\endfirsthead

\multicolumn{3}{c}%
{{\bfseries \tablename\ \thetable{} -- continued from previous page}} \\
\hline \multicolumn{1}{|c|}{\textbf{$F_{[X]}$ $S_{[X]}$}} & \multicolumn{1}{c|}{$FS_{[1111]}$ Type}
& \multicolumn{1}{c|}{Explicit wave function} \\ \hline
\endhead

\hline \multicolumn{3}{|r|}{{Continued on next page}} \\ \hline
\endfoot

\hline \hline
\endlastfoot

$F_{[211]} S_{[31]}$ & $A$ & $-\frac{1}{\sqrt{3}}
\phi_{[211]_{\rho}} \chi_{[31]_{\lambda}} +\frac{1}{\sqrt{3}}
\phi_{[211]_{\lambda}} \chi_{[31]_{\rho}}+\frac{1}{\sqrt{3}}
\phi_{[211]_{\eta}} \chi_{[31]_{\eta}}$ \\
 \hline

$F_{[22]} S_{[22]}$ & $A$ & $-\frac{1}{\sqrt{2}}
\phi_{[22]_{\rho}} \chi_{[22]_{\lambda}} +\frac{1}{\sqrt{2}}
\phi_{[22]_{\lambda}} \chi_{[22]_{\rho}}$ \\
\end{longtable}

\section{Spatial wave function of the $q^4$ subsystem}\label{sec:AP2}

In this appendix the spatial wave functions of the $q^4$ subsystem of pentaquarks with the permutation symmetries $[4]_S$ are listed in Table \ref{norp0} up to $N'=22$, where $l_\rho$, $l_\lambda$, $l_\eta$, and $L'$ are limited to $0$ and $1$ only.
Note that we have set $M'=0$ and used the abbreviation,
\begin{eqnarray}
 && \sum_{\{n_i,l_i,m_i\}} C_{n_{\rho}, l_{\rho},m_{\rho},n_{\lambda},l_{\lambda},m_{\lambda},n_{\eta},l_{\eta},m_{\eta}} \psi_{n_{\rho}l_{\rho}m_{\rho}}(\vec\rho\,)\psi_{n_{\lambda}l_{\lambda}m_{\lambda}}(\vec\lambda\,)\psi_{n_{\eta}l_{\eta}m_{\eta}}(\vec\eta\,) \nonumber \\
 &\equiv&  \sum_{\{n_i,l_i\}} C_{n_{\rho}, l_{\rho} ,n_{\lambda},l_{\lambda},n_{\eta},l_{\eta}}\,\psi(n_{\rho},l_{\rho},n_{\lambda},l_{\lambda}, n_{\eta},l_{\eta}) \nonumber \\
 &\equiv&  \sum_{\{n_i,l_i\}} C_{n_{\rho}, l_{\rho} ,n_{\lambda},l_{\lambda},n_{\eta},l_{\eta}}\,(n_{\rho},l_{\rho},n_{\lambda},l_{\lambda}, n_{\eta},l_{\eta}). \nonumber \\
 \end{eqnarray}

\begin{longtable}{|l|l|}
\caption{Normalized $q^4$ spatial wave functions with quantum number, $N'=2n$ and $L'=M'=0$} \label{norp0} \\

\hline \multicolumn{1}{|c|}{\textbf{$N'L'M'_{[X]_y}$}} & \multicolumn{1}{c|}{$C_{n_{\rho}, l_{\rho} ,n_{\lambda},l_{\lambda},n_{\eta},l_{\eta}}\psi(n_{\rho},l_{\rho},n_{\lambda},l_{\lambda}, n_{\eta},l_{\eta})$}  \\ \hline
\endfirsthead

\multicolumn{2}{c}%
{{\tablename\ \thetable{} -- continued from previous page}} \\
\hline \multicolumn{1}{|c|}{\textbf{$N'L'M'_{[X]_y}$}} &
\multicolumn{1}{c|}{$C_{n_{\rho}, l_{\rho} ,n_{\lambda},l_{\lambda},n_{\eta},l_{\eta}} \psi(n_{\rho},l_{\rho},n_{\lambda},l_{\lambda}, n_{\eta},l_{\eta})$} \\ \hline
\endhead

\hline \multicolumn{2}{|r|}{{Continued on next page}} \\ \hline
\endfoot

\hline \hline
\endlastfoot

$000_{[4]_S}$ & ($0,0,0,0,0,0$) \\ \hline
$200_{[4]_S}$ & $\frac{1}{\sqrt{3}}$($1,0,0,0,0,0$), $\frac{1}{\sqrt{3}}$($0,0,1,0,0,0$), $\frac{1}{\sqrt{3}}$($0,0,0,0,1,0$)  \\
 \hline
$400_{[4]_S}$ & $\sqrt{\frac{5}{33}}$($2,0,0,0,0,0$), $\sqrt{\frac{5}{33}}$($0,0,2,0,0,0$), $\sqrt{\frac{5}{33}}$($0,0,0,0,2,0$),  \\
\phantom{}  & $\sqrt{\frac{2}{11}}$($1,0,1,0,0,0$), $\sqrt{\frac{2}{11}}$($1,0,0,0,1,0$), $\sqrt{\frac{2}{11}}$($0,0,1,0,1,0$)  \\
 \hline
$600_{[4]_S}$ & $\sqrt{\frac{35}{429}}$($3,0,0,0,0,0$), $\sqrt{\frac{35}{429}}$($0,0,3,0,0,0$), $\sqrt{\frac{35}{429}}$($0,0,0,0,3,0$),   \\
\phantom{}  & $\sqrt{\frac{15}{143}}$($2,0,1,0,0,0$), $\sqrt{\frac{15}{143}}$($2,0,0,0,1,0$), $\sqrt{\frac{15}{143}}$($1,0,2,0,0,0$),  \\
\phantom{}  & $\sqrt{\frac{15}{143}}$($0,0,2,0,1,0$), $\sqrt{\frac{15}{143}}$($1,0,0,0,2,0$), $\sqrt{\frac{15}{143}}$($0,0,1,0,2,0$),  \\
\phantom{}  & $\sqrt{\frac{18}{143}}$($1,0,1,0,1,0$)  \\
 \hline
$800_{[4]_S}$ & $\sqrt{\frac{7}{143}}$($4,0,0,0,0,0$), $\sqrt{\frac{7}{143}}$($0,0,4,0,0,0$), $\sqrt{\frac{7}{143}}$($0,0,0,0,4,0$),   \\
\phantom{}  & $\sqrt{\frac{28}{429}}$($3,0,1,0,0,0$), $\sqrt{\frac{28}{429}}$($3,0,0,0,1,0$), $\sqrt{\frac{28}{429}}$($1,0,3,0,0,0$),  \\
\phantom{}  & $\sqrt{\frac{28}{429}}$($0,0,3,0,1,0$), $\sqrt{\frac{28}{429}}$($1,0,0,0,3,0$), $\sqrt{\frac{28}{429}}$($0,0,1,0,3,0$),  \\
\phantom{}  & $\sqrt{\frac{10}{143}}$($0,0,2,0,2,0$), $\sqrt{\frac{10}{143}}$($2,0,0,0,2,0$), $\sqrt{\frac{10}{143}}$($2,0,2,0,0,0$),  \\
\phantom{}  & $\sqrt{\frac{12}{143}}$($2,0,1,0,1,0$), $\sqrt{\frac{12}{143}}$($1,0,2,0,1,0$), $\sqrt{\frac{12}{143}}$($1,0,1,0,2,0$)  \\
 \hline
 $1000_{[4]_S}$ & $\sqrt{\frac{7}{221}}$($5,0,0,0,0,0$), $\sqrt{\frac{7}{221}}$($0,0,5,0,0,0$), $\sqrt{\frac{7}{221}}$($0,0,0,0,5,0$),   \\
\phantom{}  & $\sqrt{\frac{105}{2431}}$($4,0,1,0,0,0$), $\sqrt{\frac{105}{2431}}$($4,0,0,0,1,0$), $\sqrt{\frac{105}{2431}}$($1,0,4,0,0,0$),  \\
\phantom{}  & $\sqrt{\frac{105}{2431}}$($0,0,4,0,1,0$), $\sqrt{\frac{105}{2431}}$($1,0,0,0,4,0$), $\sqrt{\frac{105}{2431}}$($0,0,1,0,4,0$),  \\
\phantom{}  & $\sqrt{\frac{350}{7293}}$($3,0,2,0,0,0$), $\sqrt{\frac{350}{7293}}$($3,0,0,0,2,0$), $\sqrt{\frac{350}{7293}}$($2,0,3,0,0,0$),  \\
\phantom{}  & $\sqrt{\frac{350}{7293}}$($0,0,3,0,2,0$), $\sqrt{\frac{350}{7293}}$($2,0,0,0,3,0$), $\sqrt{\frac{350}{7293}}$($0,0,2,0,3,0$),  \\
\phantom{}  & $\sqrt{\frac{150}{2431}}$($1,0,2,0,2,0$), $\sqrt{\frac{150}{2431}}$($2,0,1,0,2,0$), $\sqrt{\frac{150}{2431}}$($2,0,2,0,1,0$),  \\
\phantom{}  & $\sqrt{\frac{140}{2431}}$($3,0,1,0,1,0$), $\sqrt{\frac{140}{2431}}$($1,0,3,0,1,0$), $\sqrt{\frac{140}{2431}}$($1,0,1,0,3,0$)  \\
\hline
$1200_{[4]_S}$ & $\sqrt{\frac{7}{323}}$($6,0,0,0,0,0$), $\sqrt{\frac{7}{323}}$($0,0,6,0,0,0$), $\sqrt{\frac{7}{323}}$($0,0,0,0,6,0$),   \\
\phantom{}  & $\sqrt{\frac{126}{4199}}$($5,0,1,0,0,0$), $\sqrt{\frac{126}{4199}}$($5,0,0,0,1,0$), $\sqrt{\frac{126}{4199}}$($1,0,5,0,0,0$),  \\
\phantom{}  & $\sqrt{\frac{126}{4199}}$($0,0,5,0,1,0$), $\sqrt{\frac{126}{4199}}$($1,0,0,0,5,0$), $\sqrt{\frac{126}{4199}}$($0,0,1,0,5,0$),  \\
\phantom{}  & $\sqrt{\frac{1575}{46189}}$($4,0,2,0,0,0$), $\sqrt{\frac{1575}{46189}}$($4,0,0,0,2,0$), $\sqrt{\frac{1575}{46189}}$($2,0,4,0,0,0$),  \\
\phantom{}  & $\sqrt{\frac{1575}{46189}}$($0,0,4,0,2,0$), $\sqrt{\frac{1575}{46189}}$($2,0,0,0,4,0$), $\sqrt{\frac{1575}{46189}}$($0,0,2,0,4,0$),  \\
\phantom{}  & $\frac{70}{\sqrt{138567}}$($0,0,3,0,3,0$), $\frac{70}{\sqrt{138567}}$($3,0,0,0,3,0$), $\frac{70}{\sqrt{138567}}$($3,0,3,0,0,0$),  \\
\phantom{}  & $\sqrt{\frac{2100}{46189}}$($3,0,1,0,2,0$), $\sqrt{\frac{2100}{46189}}$($3,0,2,0,1,0$), $\sqrt{\frac{2100}{46189}}$($1,0,3,0,2,0$),  \\
\phantom{}  & $\sqrt{\frac{2100}{46189}}$($2,0,3,0,1,0$), $\sqrt{\frac{2100}{46189}}$($1,0,2,0,3,0$), $\sqrt{\frac{2100}{46189}}$($2,0,1,0,3,0$),  \\
\phantom{}  & $\sqrt{\frac{1890}{46189}}$($4,0,1,0,1,0$), $\sqrt{\frac{1890}{46189}}$($1,0,4,0,1,0$), $\sqrt{\frac{1890}{46189}}$($1,0,1,0,4,0$),  \\
\phantom{}  & $\sqrt{\frac{2250}{46189}}$($2,0,2,0,2,0$)  \\
\hline

 $1400_{[4]_S}$ & $\sqrt{\frac{5}{323}}$($7,0,0,0,0,0$), $\sqrt{\frac{5}{323}}$($0,0,7,0,0,0$), $\sqrt{\frac{5}{323}}$($0,0,0,0,7,0$),   \\
\phantom{}  & $\sqrt{\frac{7}{323}}$($6,0,1,0,0,0$), $\sqrt{\frac{7}{323}}$($6,0,0,0,1,0$), $\sqrt{\frac{7}{323}}$($1,0,6,0,0,0$),  \\
\phantom{}  & $\sqrt{\frac{7}{323}}$($0,0,6,0,1,0$), $\sqrt{\frac{7}{323}}$($1,0,0,0,6,0$), $\sqrt{\frac{7}{323}}$($0,0,1,0,6,0$),  \\
\phantom{}  & $\sqrt{\frac{105}{4199}}$($5,0,2,0,0,0$), $\sqrt{\frac{105}{4199}}$($5,0,0,0,2,0$), $\sqrt{\frac{105}{4199}}$($2,0,5,0,0,0$),  \\
\phantom{}  & $\sqrt{\frac{105}{4199}}$($0,0,5,0,2,0$), $\sqrt{\frac{105}{4199}}$($2,0,0,0,5,0$), $\sqrt{\frac{105}{4199}}$($0,0,2,0,5,0$),  \\
\phantom{}  & $\sqrt{\frac{126}{4199}}$($5,0,1,0,1,0$), $\sqrt{\frac{126}{4199}}$($1,0,5,0,1,0$), $\sqrt{\frac{126}{4199}}$($1,0,1,0,5,0$),  \\
\phantom{}  & $\frac{35}{\sqrt{46189}}$($0,0,4,0,3,0$), $\frac{35}{\sqrt{46189}}$($0,0,3,0,4,0$), $\frac{35}{\sqrt{46189}}$($4,0,0,0,3,0$),  \\
\phantom{}  & $\frac{35}{\sqrt{46189}}$($3,0,0,0,4,0$), $\frac{35}{\sqrt{46189}}$($4,0,3,0,0,0$), $\frac{35}{\sqrt{46189}}$($3,0,4,0,0,0$),  \\
\phantom{}  & $\sqrt{\frac{1575}{46189}}$($4,0,1,0,2,0$), $\sqrt{\frac{1575}{46189}}$($4,0,2,0,1,0$), $\sqrt{\frac{1575}{46189}}$($1,0,4,0,2,0$),  \\
\phantom{}  & $\sqrt{\frac{1575}{46189}}$($2,0,4,0,1,0$), $\sqrt{\frac{1575}{46189}}$($1,0,2,0,4,0$), $\sqrt{\frac{1575}{46189}}$($2,0,1,0,4,0$),  \\
\phantom{}  & $\frac{70}{\sqrt{138567}}$($1,0,3,0,3,0$), $\frac{70}{\sqrt{138567}}$($3,0,1,0,3,0$), $\frac{70}{\sqrt{138567}}$($3,0,3,0,1,0$),  \\
\phantom{}  & $\sqrt{\frac{1750}{46189}}$($3,0,2,0,2,0$), $\sqrt{\frac{1750}{46189}}$($2,0,3,0,2,0$), $\sqrt{\frac{1750}{46189}}$($2,0,2,0,3,0$)  \\
\hline

$1600_{[4]_S}$ & $\sqrt{\frac{5}{437}}$($8,0,0,0,0,0$), $\sqrt{\frac{5}{437}}$($0,0,8,0,0,0$), $\sqrt{\frac{5}{437}}$($0,0,0,0,8,0$),   \\
\phantom{}  & $\sqrt{\frac{120}{7429}}$($7,0,1,0,0,0$), $\sqrt{\frac{120}{7429}}$($7,0,0,0,1,0$), $\sqrt{\frac{120}{7429}}$($1,0,7,0,0,0$),  \\
\phantom{}  & $\sqrt{\frac{120}{7429}}$($0,0,7,0,1,0$), $\sqrt{\frac{120}{7429}}$($1,0,0,0,7,0$), $\sqrt{\frac{120}{7429}}$($0,0,1,0,7,0$),  \\
\phantom{}  & $\sqrt{\frac{140}{7429}}$($6,0,2,0,0,0$), $\sqrt{\frac{140}{7429}}$($6,0,0,0,2,0$), $\sqrt{\frac{140}{7429}}$($2,0,6,0,0,0$),  \\
\phantom{}  & $\sqrt{\frac{140}{7429}}$($0,0,6,0,2,0$), $\sqrt{\frac{140}{7429}}$($2,0,0,0,6,0$), $\sqrt{\frac{140}{7429}}$($0,0,2,0,6,0$),  \\
\phantom{}  & $\sqrt{\frac{168}{7429}}$($6,0,1,0,1,0$), $\sqrt{\frac{168}{7429}}$($1,0,6,0,1,0$), $\sqrt{\frac{168}{7429}}$($1,0,1,0,6,0$),  \\
\phantom{}  & $\sqrt{\frac{1960}{96577}}$($0,0,5,0,3,0$), $\sqrt{\frac{1960}{96577}}$($0,0,3,0,5,0$), $\sqrt{\frac{1960}{96577}}$($5,0,0,0,3,0$),  \\
\phantom{}  & $\sqrt{\frac{1960}{96577}}$($3,0,0,0,5,0$), $\sqrt{\frac{1960}{96577}}$($5,0,3,0,0,0$), $\sqrt{\frac{1960}{96577}}$($3,0,5,0,0,0$),  \\
\phantom{}  & $\sqrt{\frac{2520}{96577}}$($5,0,1,0,2,0$), $\sqrt{\frac{2520}{96577}}$($5,0,2,0,1,0$), $\sqrt{\frac{2520}{96577}}$($1,0,5,0,2,0$),  \\
\phantom{}  & $\sqrt{\frac{2520}{96577}}$($2,0,5,0,1,0$), $\sqrt{\frac{2520}{96577}}$($1,0,2,0,5,0$), $\sqrt{\frac{2520}{96577}}$($2,0,1,0,5,0$),  \\
\phantom{}  & $\sqrt{\frac{22050}{1062347}}$($0,0,4,0,4,0$), $\sqrt{\frac{22050}{1062347}}$($4,0,0,0,4,0$), $\sqrt{\frac{22050}{1062347}}$($4,0,4,0,0,0$),  \\
\phantom{}  & $\sqrt{\frac{29400}{1062347}}$($4,0,1,0,3,0$), $\sqrt{\frac{29400}{1062347}}$($4,0,3,0,1,0$), $\sqrt{\frac{29400}{1062347}}$($1,0,4,0,3,0$),  \\
\phantom{}  & $\sqrt{\frac{29400}{1062347}}$($3,0,4,0,1,0$), $\sqrt{\frac{29400}{1062347}}$($1,0,3,0,4,0$), $\sqrt{\frac{29400}{1062347}}$($3,0,1,0,4,0$),  \\
\phantom{}  & $\sqrt{\frac{31500}{1062347}}$($4,0,2,0,2,0$), $\sqrt{\frac{31500}{1062347}}$($2,0,4,0,2,0$), $\sqrt{\frac{31500}{1062347}}$($2,0,2,0,4,0$),  \\
\phantom{}  & $\sqrt{\frac{98000}{3187041}}$($2,0,3,0,3,0$), $\sqrt{\frac{98000}{3187041}}$($3,0,2,0,3,0$), $\sqrt{\frac{98000}{3187041}}$($3,0,3,0,2,0$)  \\
\hline

$1800_{[4]_S}$ & $\frac{1}{\sqrt{115}}$($9,0,0,0,0,0$), $\frac{1}{\sqrt{115}}$($0,0,9,0,0,0$), $\frac{1}{\sqrt{115}}$($0,0,0,0,9,0$),   \\
\phantom{}  & $\sqrt{\frac{27}{2185}}$($8,0,1,0,0,0$), $\sqrt{\frac{27}{2185}}$($8,0,0,0,1,0$), $\sqrt{\frac{27}{2185}}$($1,0,8,0,0,0$),  \\
\phantom{}  & $\sqrt{\frac{27}{2185}}$($0,0,8,0,1,0$), $\sqrt{\frac{27}{2185}}$($1,0,0,0,8,0$), $\sqrt{\frac{27}{2185}}$($0,0,1,0,8,0$),  \\
\phantom{}  & $\sqrt{\frac{108}{7429}}$($7,0,2,0,0,0$), $\sqrt{\frac{108}{7429}}$($7,0,0,0,2,0$), $\sqrt{\frac{108}{7429}}$($2,0,7,0,0,0$),  \\
\phantom{}  & $\sqrt{\frac{108}{7429}}$($0,0,7,0,2,0$), $\sqrt{\frac{108}{7429}}$($2,0,0,0,7,0$), $\sqrt{\frac{108}{7429}}$($0,0,2,0,7,0$),  \\
\phantom{}  & $\sqrt{\frac{648}{37145}}$($7,0,1,0,1,0$), $\sqrt{\frac{648}{37145}}$($1,0,7,0,1,0$), $\sqrt{\frac{648}{37145}}$($1,0,1,0,7,0$),  \\
\phantom{}  & $\sqrt{\frac{588}{37145}}$($0,0,6,0,3,0$), $\sqrt{\frac{588}{37145}}$($0,0,3,0,6,0$), $\sqrt{\frac{588}{37145}}$($6,0,0,0,3,0$),  \\
\phantom{}  & $\sqrt{\frac{588}{37145}}$($3,0,0,0,6,0$), $\sqrt{\frac{588}{37145}}$($6,0,3,0,0,0$), $\sqrt{\frac{588}{37145}}$($3,0,6,0,0,0$),  \\
\phantom{}  & $\sqrt{\frac{756}{37145}}$($6,0,1,0,2,0$), $\sqrt{\frac{756}{37145}}$($6,0,2,0,1,0$), $\sqrt{\frac{756}{37145}}$($1,0,6,0,2,0$),  \\
\phantom{}  & $\sqrt{\frac{756}{37145}}$($2,0,6,0,1,0$), $\sqrt{\frac{756}{37145}}$($1,0,2,0,6,0$), $\sqrt{\frac{756}{37145}}$($2,0,1,0,6,0$),  \\
\phantom{}  & $\sqrt{\frac{7938}{482885}}$($0,0,5,0,4,0$), $\sqrt{\frac{7938}{482885}}$($0,0,4,0,5,0$), $\sqrt{\frac{7938}{482885}}$($5,0,0,0,4,0$),  \\
\phantom{}  & $\sqrt{\frac{7938}{482885}}$($4,0,0,0,5,0$), $\sqrt{\frac{7938}{482885}}$($5,0,4,0,0,0$), $\sqrt{\frac{7938}{482885}}$($4,0,5,0,0,0$),  \\
\phantom{}  & $\sqrt{\frac{10584}{482885}}$($5,0,1,0,3,0$), $\sqrt{\frac{10584}{482885}}$($5,0,3,0,1,0$), $\sqrt{\frac{10584}{482885}}$($1,0,5,0,3,0$),  \\
\phantom{}  & $\sqrt{\frac{10584}{482885}}$($3,0,5,0,1,0$), $\sqrt{\frac{10584}{482885}}$($1,0,3,0,5,0$), $\sqrt{\frac{10584}{482885}}$($3,0,1,0,5,0$),  \\
\phantom{}  & $\sqrt{\frac{2268}{96577}}$($5,0,2,0,2,0$), $\sqrt{\frac{2268}{96577}}$($2,0,5,0,2,0$), $\sqrt{\frac{2268}{96577}}$($2,0,2,0,5,0$),  \\
\phantom{}  & $\sqrt{\frac{23814}{1062347}}$($1,0,4,0,4,0$), $\sqrt{\frac{23814}{1062347}}$($4,0,1,0,4,0$), $\sqrt{\frac{23814}{1062347}}$($4,0,4,0,1,0$),  \\
\phantom{}  & $\sqrt{\frac{26460}{1062347}}$($4,0,2,0,3,0$), $\sqrt{\frac{26460}{1062347}}$($4,0,3,0,2,0$), $\sqrt{\frac{26460}{1062347}}$($2,0,4,0,3,0$),  \\
\phantom{}  & $\sqrt{\frac{26460}{1062347}}$($3,0,4,0,2,0$), $\sqrt{\frac{26460}{1062347}}$($2,0,3,0,4,0$), $\sqrt{\frac{26460}{1062347}}$($3,0,2,0,4,0$),  \\
\phantom{}  & $\sqrt{\frac{27440}{1062347}}$($3,0,3,0,3,0$)  \\
\hline
$2000_{[4]_S}$ & $\sqrt{\frac{7}{1035}}$($10,0,0,0,0,0$), $\sqrt{\frac{7}{1035}}$($0,0,10,0,0,0$), $\sqrt{\frac{7}{1035}}$($0,0,0,0,10,0$),   \\
\phantom{}  & $\sqrt{\frac{2}{207}}$($9,0,1,0,0,0$), $\sqrt{\frac{2}{207}}$($9,0,0,0,1,0$), $\sqrt{\frac{2}{207}}$($1,0,9,0,0,0$),  \\
\phantom{}  & $\sqrt{\frac{2}{207}}$($0,0,9,0,1,0$), $\sqrt{\frac{2}{207}}$($1,0,0,0,9,0$), $\sqrt{\frac{2}{207}}$($0,0,1,0,9,0$),  \\
\phantom{}  & $\sqrt{\frac{5}{437}}$($8,0,2,0,0,0$), $\sqrt{\frac{5}{437}}$($8,0,0,0,2,0$), $\sqrt{\frac{5}{437}}$($2,0,8,0,0,0$),  \\
\phantom{}  & $\sqrt{\frac{5}{437}}$($0,0,8,0,2,0$), $\sqrt{\frac{5}{437}}$($2,0,0,0,8,0$), $\sqrt{\frac{5}{437}}$($0,0,2,0,8,0$),  \\
\phantom{}  & $\sqrt{\frac{6}{437}}$($8,0,1,0,1,0$), $\sqrt{\frac{6}{437}}$($1,0,8,0,1,0$), $\sqrt{\frac{6}{437}}$($1,0,1,0,8,0$),  \\
\phantom{}  & $\sqrt{\frac{280}{22287}}$($0,0,7,0,3,0$), $\sqrt{\frac{280}{22287}}$($0,0,3,0,7,0$), $\sqrt{\frac{280}{22287}}$($7,0,0,0,3,0$),  \\
\phantom{}  & $\sqrt{\frac{280}{22287}}$($3,0,0,0,7,0$), $\sqrt{\frac{280}{22287}}$($7,0,3,0,0,0$), $\sqrt{\frac{280}{22287}}$($3,0,7,0,0,0$),  \\
\phantom{}  & $\sqrt{\frac{120}{7429}}$($7,0,1,0,2,0$), $\sqrt{\frac{120}{7429}}$($7,0,2,0,1,0$), $\sqrt{\frac{120}{7429}}$($1,0,7,0,2,0$),  \\
\phantom{}  & $\sqrt{\frac{120}{7429}}$($2,0,7,0,1,0$), $\sqrt{\frac{120}{7429}}$($1,0,2,0,7,0$), $\sqrt{\frac{120}{7429}}$($2,0,1,0,7,0$),  \\
\phantom{}  & $\sqrt{\frac{98}{7429}}$($0,0,6,0,4,0$), $\sqrt{\frac{98}{7429}}$($0,0,4,0,6,0$), $\sqrt{\frac{98}{7429}}$($6,0,0,0,4,0$),  \\
\phantom{}  & $\sqrt{\frac{98}{7429}}$($4,0,0,0,6,0$), $\sqrt{\frac{98}{7429}}$($6,0,4,0,0,0$), $\sqrt{\frac{98}{7429}}$($4,0,6,0,0,0$),  \\
\phantom{}  & $\sqrt{\frac{392}{22287}}$($6,0,1,0,3,0$), $\sqrt{\frac{392}{22287}}$($6,0,3,0,1,0$), $\sqrt{\frac{392}{22287}}$($1,0,6,0,3,0$),  \\
\phantom{}  & $\sqrt{\frac{392}{22287}}$($3,0,6,0,1,0$), $\sqrt{\frac{392}{22287}}$($1,0,3,0,6,0$), $\sqrt{\frac{392}{22287}}$($3,0,1,0,6,0$),  \\
\phantom{}  & $\sqrt{\frac{140}{7429}}$($6,0,2,0,2,0$), $\sqrt{\frac{140}{7429}}$($2,0,6,0,2,0$), $\sqrt{\frac{140}{7429}}$($2,0,2,0,6,0$),  \\
\phantom{}  & $\sqrt{\frac{6468}{482885}}$($0,0,5,0,5,0$), $\sqrt{\frac{6468}{482885}}$($5,0,0,0,5,0$), $\sqrt{\frac{6468}{482885}}$($5,0,5,0,0,0$),  \\
\phantom{}  & $\frac{42}{\sqrt{96577}}$($5,0,1,0,4,0$), $\frac{42}{\sqrt{96577}}$($5,0,4,0,1,0$), $\frac{42}{\sqrt{96577}}$($1,0,5,0,4,0$),  \\
\phantom{}  & $\frac{42}{\sqrt{96577}}$($4,0,5,0,1,0$), $\frac{42}{\sqrt{96577}}$($1,0,4,0,5,0$), $\frac{42}{\sqrt{96577}}$($4,0,1,0,5,0$),  \\
\phantom{}  & $\sqrt{\frac{1960}{96577}}$($5,0,2,0,3,0$), $\sqrt{\frac{1960}{96577}}$($5,0,3,0,2,0$), $\sqrt{\frac{1960}{96577}}$($2,0,5,0,3,0$),  \\
\phantom{}  & $\sqrt{\frac{1960}{96577}}$($3,0,5,0,2,0$), $\sqrt{\frac{1960}{96577}}$($2,0,3,0,5,0$), $\sqrt{\frac{1960}{96577}}$($3,0,2,0,5,0$),  \\
\phantom{}  & $\sqrt{\frac{22050}{1062347}}$($2,0,4,0,4,0$), $\sqrt{\frac{22050}{1062347}}$($4,0,2,0,4,0$), $\sqrt{\frac{22050}{1062347}}$($4,0,4,0,2,0$),  \\
\phantom{}  & $\sqrt{\frac{68600}{3187041}}$($4,0,3,0,3,0$), $\sqrt{\frac{68600}{3187041}}$($3,0,4,0,3,0$), $\sqrt{\frac{68600}{3187041}}$($3,0,3,0,4,0$)  \\
\hline

$2200_{[4]_S}$ & $\sqrt{\frac{7}{1305}}$($11,0,0,0,0,0$), $\sqrt{\frac{7}{1305}}$($0,0,11,0,0,0$), $\sqrt{\frac{7}{1305}}$($0,0,0,0,11,0$),   \\
\phantom{}  & $\sqrt{\frac{77}{10005}}$($10,0,1,0,0,0$), $\sqrt{\frac{77}{10005}}$($10,0,0,0,1,0$), $\sqrt{\frac{77}{10005}}$($1,0,10,0,0,0$),  \\
\phantom{}  & $\sqrt{\frac{77}{10005}}$($0,0,10,0,1,0$), $\sqrt{\frac{77}{10005}}$($1,0,0,0,10,0$), $\sqrt{\frac{77}{10005}}$($0,0,1,0,10,0$),  \\
\phantom{}  & $\sqrt{\frac{55}{6003}}$($9,0,2,0,0,0$), $\sqrt{\frac{55}{6003}}$($9,0,0,0,2,0$), $\sqrt{\frac{55}{6003}}$($2,0,9,0,0,0$),  \\
\phantom{}  & $\sqrt{\frac{55}{6003}}$($0,0,9,0,2,0$), $\sqrt{\frac{55}{6003}}$($2,0,0,0,9,0$), $\sqrt{\frac{55}{6003}}$($0,0,2,0,9,0$),  \\
\phantom{}  & $\sqrt{\frac{22}{2001}}$($9,0,1,0,1,0$), $\sqrt{\frac{22}{2001}}$($1,0,9,0,1,0$), $\sqrt{\frac{22}{2001}}$($1,0,1,0,9,0$),  \\
\phantom{}  & $\sqrt{\frac{385}{38019}}$($0,0,8,0,3,0$), $\sqrt{\frac{385}{38019}}$($0,0,3,0,8,0$), $\sqrt{\frac{385}{38019}}$($8,0,0,0,3,0$),  \\
\phantom{}  & $\sqrt{\frac{385}{38019}}$($3,0,0,0,8,0$), $\sqrt{\frac{385}{38019}}$($8,0,3,0,0,0$), $\sqrt{\frac{385}{38019}}$($3,0,8,0,0,0$),  \\
\phantom{}  & $\sqrt{\frac{165}{12673}}$($8,0,1,0,2,0$), $\sqrt{\frac{165}{12673}}$($8,0,2,0,1,0$), $\sqrt{\frac{165}{12673}}$($1,0,8,0,2,0$),  \\
\phantom{}  & $\sqrt{\frac{165}{12673}}$($2,0,8,0,1,0$), $\sqrt{\frac{165}{12673}}$($1,0,2,0,8,0$), $\sqrt{\frac{165}{12673}}$($2,0,1,0,8,0$),  \\
\phantom{}  & $\sqrt{\frac{2310}{215441}}$($0,0,7,0,4,0$), $\sqrt{\frac{2310}{215441}}$($0,0,4,0,7,0$), $\sqrt{\frac{2310}{215441}}$($7,0,0,0,4,0$),  \\
\phantom{}  & $\sqrt{\frac{2310}{215441}}$($4,0,0,0,7,0$), $\sqrt{\frac{2310}{215441}}$($7,0,4,0,0,0$), $\sqrt{\frac{2310}{215441}}$($4,0,7,0,0,0$),  \\
\phantom{}  & $\sqrt{\frac{3080}{215441}}$($7,0,1,0,3,0$), $\sqrt{\frac{3080}{215441}}$($7,0,3,0,1,0$), $\sqrt{\frac{3080}{215441}}$($1,0,7,0,3,0$),  \\
\phantom{}  & $\sqrt{\frac{3080}{215441}}$($3,0,7,0,1,0$), $\sqrt{\frac{3080}{215441}}$($1,0,3,0,7,0$), $\sqrt{\frac{3080}{215441}}$($3,0,1,0,7,0$),  \\
\phantom{}  & $\sqrt{\frac{3300}{215441}}$($7,0,2,0,2,0$), $\sqrt{\frac{3300}{215441}}$($2,0,7,0,2,0$), $\sqrt{\frac{3300}{215441}}$($2,0,2,0,7,0$),  \\
\phantom{}  & $\sqrt{\frac{11858}{1077205}}$($0,0,5,0,6,0$), $\sqrt{\frac{11858}{1077205}}$($0,0,6,0,5,0$), $\sqrt{\frac{11858}{1077205}}$($5,0,0,0,6,0$),  \\
\phantom{}  & $\sqrt{\frac{11858}{1077205}}$($6,0,0,0,5,0$), $\sqrt{\frac{11858}{1077205}}$($5,0,6,0,0,0$), $\sqrt{\frac{11858}{1077205}}$($6,0,5,0,0,0$),  \\
\phantom{}  & $\sqrt{\frac{3234}{215441}}$($6,0,1,0,4,0$), $\sqrt{\frac{3234}{215441}}$($6,0,4,0,1,0$), $\sqrt{\frac{3234}{215441}}$($1,0,6,0,4,0$),  \\
\phantom{}  & $\sqrt{\frac{3234}{215441}}$($4,0,6,0,1,0$), $\sqrt{\frac{3234}{215441}}$($1,0,4,0,6,0$), $\sqrt{\frac{3234}{215441}}$($4,0,1,0,6,0$),  \\
\phantom{}  & $\sqrt{\frac{10780}{646323}}$($6,0,2,0,3,0$), $\sqrt{\frac{10780}{646323}}$($6,0,3,0,2,0$), $\sqrt{\frac{10780}{646323}}$($2,0,6,0,3,0$),  \\
\phantom{}  & $\sqrt{\frac{10780}{646323}}$($3,0,6,0,2,0$), $\sqrt{\frac{10780}{646323}}$($2,0,3,0,6,0$), $\sqrt{\frac{10780}{646323}}$($3,0,2,0,6,0$),  \\
\phantom{}  & $\frac{462}{\sqrt{14003665}}$($1,0,5,0,5,0$), $\frac{462}{\sqrt{14003665}}$($5,0,1,0,5,0$), $\frac{462}{\sqrt{14003665}}$($5,0,5,0,1,0$),  \\
\phantom{}  & $\sqrt{\frac{48510}{2800733}}$($5,0,2,0,4,0$), $\sqrt{\frac{48510}{2800733}}$($5,0,4,0,2,0$), $\sqrt{\frac{48510}{2800733}}$($2,0,5,0,4,0$),  \\
\phantom{}  & $\sqrt{\frac{48510}{2800733}}$($4,0,5,0,2,0$), $\sqrt{\frac{48510}{2800733}}$($2,0,4,0,5,0$), $\sqrt{\frac{48510}{2800733}}$($4,0,2,0,5,0$),  \\
\phantom{}  & $\sqrt{\frac{150920}{8402199}}$($5,0,3,0,3,0$), $\sqrt{\frac{150920}{8402199}}$($3,0,5,0,3,0$), $\sqrt{\frac{150920}{8402199}}$($3,0,3,0,5,0$),  \\
\phantom{}  & $\sqrt{\frac{51450}{2800733}}$($3,0,4,0,4,0$), $\sqrt{\frac{51450}{2800733}}$($4,0,3,0,4,0$), $\sqrt{\frac{51450}{2800733}}$($4,0,4,0,3,0$)  \\
\end{longtable}

\section{Spatial wave function of the $q^4 \overline q$ system}\label{sec:AP3}

In this appendix the spatial wave functions of pentaquarks with the $q^4$ symmetry $[4]_S$ are shown in Table \ref{norp4}, where $\psi^{q^{4}}_{N'L'M'}$ ($L'=M'=0$) and
$\psi_{n_\xi,l_\xi}(\vec\xi\,)$ ($l_\xi=0$) are
the spatial wave functions of the $q^4$ subsystem and the harmonic oscillator wave function for the $\vec\xi$ coordinate, respectively.
With the limitation $n_\xi\leq 4$, one may have up to five degenerate states for each pentaquark energy level.
\begin{center}
\begin{table}[h!b!p!]
\caption[Pentaquark spatial wave functions of symmetric type]{Pentaquark spatial wave functions of symmetric type.}
\begin{tabular}{|c|c|}
\hline
$\Psi^{q^{4}\overline q}_{000_{[4]_S}}$ & $\psi^{q^4}_{000_{[4]_S}}\psi_{0,0}(\vec\xi\,)$ \\
\hline
$\Psi^{q^{4}\overline q}_{200_{[4]_S}}$ & $\psi^{q^4}_{200_{[4]_S}}\psi_{0,0}(\vec\xi\,)$, $\psi^{q^4}_{000_{[4]_S}}\psi_{1,0}(\vec\xi\,)$ \\
\hline
$\Psi^{q^{4}\overline q}_{400_{[4]_S}}$ & $\psi^{q^4}_{400_{[4]_S}}\psi_{0,0}(\vec\xi\,)$, $\psi^{q^4}_{200_{[4]_S}}\psi_{1,0}(\vec\xi\,)$, $\psi^{q^4}_{000_{[4]_S}}\psi_{2,0}(\vec\xi\,)$ \\
\hline
$\Psi^{q^{4}\overline q}_{600_{[4]_S}}$ & $\psi^{q^4}_{600_{[4]_S}}\psi_{0,0}(\vec\xi)$, $\psi^{q^4}_{400_{[4]_S}}\psi_{1,0}(\vec\xi)$, $\psi^{q^4}_{200_{[4]_S}}\psi_{2,0}(\vec\xi)$, $\psi^{q^4}_{000_{[4]_S}}\psi_{3,0}(\vec\xi)$ \\
\hline
$\Psi^{q^{4}\overline q}_{800_{[4]_S}}$ & $\psi^{q^4}_{800_{[4]_S}}\psi_{0,0}(\vec\xi)$, $\psi^{q^4}_{600_{[4]_S}}\psi_{1,0}(\vec\xi)$, $\psi^{q^4}_{400_{[4]_S}}\psi_{2,0}(\vec\xi)$, $\psi^{q^4}_{200_{[4]_S}}\psi_{3,0}(\vec\xi)$, $\psi^{q^4}_{000_{[4]_S}}\psi_{4,0}(\vec\xi)$\\
\hline
$\Psi^{q^{4}\overline q}_{1000_{[4]_S}}$ & $\psi^{q^4}_{1000_{[4]_S}}\psi_{0,0}(\vec\xi)$, $\psi^{q^4}_{800_{[4]_S}}\psi_{1,0}(\vec\xi)$, $\psi^{q^4}_{600_{[4]_S}}\psi_{2,0}(\vec\xi)$, $\psi^{q^4}_{400_{[4]_S}}\psi_{3,0}(\vec\xi)$, $\psi^{q^4}_{200_{[4]_S}}\psi_{4,0}(\vec\xi)$\\
\hline
$\Psi^{q^{4}\overline q}_{1200_{[4]_S}}$ & $\psi^{q^4}_{1200_{[4]_S}}\psi_{0,0}(\vec\xi)$, $\psi^{q^4}_{1000_{[4]_S}}\psi_{1,0}(\vec\xi)$, $\psi^{q^4}_{800_{[4]_S}}\psi_{2,0}(\vec\xi)$, $\psi^{q^4}_{600_{[4]_S}}\psi_{3,0}(\vec\xi)$, $\psi^{q^4}_{400_{[4]_S}}\psi_{4,0}(\vec\xi)$\\
\hline
$\Psi^{q^{4}\overline q}_{1400_{[4]_S}}$ & $\psi^{q^4}_{1400_{[4]_S}}\psi_{0,0}(\vec\xi)$, $\psi^{q^4}_{1200_{[4]_S}}\psi_{1,0}(\vec\xi)$, $\psi^{q^4}_{1000_{[4]_S}}\psi_{2,0}(\vec\xi)$, $\psi^{q^4}_{800_{[4]_S}}\psi_{3,0}(\vec\xi)$, $\psi^{q^4}_{600_{[4]_S}}\psi_{4,0}(\vec\xi)$\\
\hline
$\Psi^{q^{4}\overline q}_{1600_{[4]_S}}$ & $\psi^{q^4}_{1600_{[4]_S}}\psi_{0,0}(\vec\xi)$, $\psi^{q^4}_{1400_{[4]_S}}\psi_{1,0}(\vec\xi)$, $\psi^{q^4}_{1200_{[4]_S}}\psi_{2,0}(\vec\xi)$, $\psi^{q^4}_{1000_{[4]_S}}\psi_{3,0}(\vec\xi)$, $\psi^{q^4}_{800_{[4]_S}}\psi_{4,0}(\vec\xi)$\\
\hline
$\Psi^{q^{4}\overline q}_{1800_{[4]_S}}$ & $\psi^{q^4}_{1800_{[4]_S}}\psi_{0,0}(\vec\xi)$, $\psi^{q^4}_{1600_{[4]_S}}\psi_{1,0}(\vec\xi)$, $\psi^{q^4}_{1400_{[4]_S}}\psi_{2,0}(\vec\xi)$, $\psi^{q^4}_{1200_{[4]_S}}\psi_{3,0}(\vec\xi)$, $\psi^{q^4}_{1000_{[4]_S}}\psi_{4,0}(\vec\xi)$\\
\hline
$\Psi^{q^{4}\overline q}_{2000_{[4]_S}}$ & $\psi^{q^4}_{2000_{[4]_S}}\psi_{0,0}(\vec\xi)$, $\psi^{q^4}_{1800_{[4]_S}}\psi_{1,0}(\vec\xi)$, $\psi^{q^4}_{1600_{[4]_S}}\psi_{2,0}(\vec\xi)$, $\psi^{q^4}_{1400_{[4]_S}}\psi_{3,0}(\vec\xi)$, $\psi^{q^4}_{1200_{[4]_S}}\psi_{4,0}(\vec\xi)$\\
\hline
$\Psi^{q^{4}\overline q}_{2200_{[4]_S}}$ & $\psi^{q^4}_{2200_{[4]_S}}\psi_{0,0}(\vec\xi)$, $\psi^{q^4}_{2000_{[4]_S}}\psi_{1,0}(\vec\xi)$, $\psi^{q^4}_{1800_{[4]_S}}\psi_{2,0}(\vec\xi)$, $\psi^{q^4}_{1600_{[4]_S}}\psi_{3,0}(\vec\xi)$, $\psi^{q^4}_{1400_{[4]_S}}\psi_{4,0}(\vec\xi)$\\\hline
\end{tabular}
\label{norp4}
\end{table}
\end{center}
\end{widetext}

\bibliographystyle{unsrt}

\end{document}